\documentclass[aps,prl,twocolumn,showpacs,superscriptaddress,footinbib]{revtex4-1} 

\usepackage[english]{babel}
\usepackage{xcolor}
\usepackage{comment}
\usepackage{graphicx}
\usepackage{stackengine}
\usepackage{dcolumn}
\usepackage{bm}
\usepackage{bbm}
\usepackage[utf8]{inputenc}
\usepackage{amssymb}
\usepackage{amsmath}
\usepackage{bbold}
\usepackage{pstricks}
\usepackage{slashed}
\usepackage{braket}
\usepackage{hyperref}
\usepackage{natbib}
\usepackage{float}
\usepackage{verbatim}
\usepackage{siunitx}
\usepackage{lipsum}
\usepackage{slashed}
\usepackage{ulem}
\usepackage[left=1.85cm,right=1.85cm,top=1.85cm,bottom=1.85cm]{geometry}

\definecolor{mygreen}{rgb}{0,0.5,0}
\definecolor{myblue}{rgb}{0,0,0.75}
\definecolor{mymagenta}{cmyk}{0,1,0,0.12}
\definecolor{mygray}{rgb}{0.5,0.5,0.5}

\newcommand{\Fig}[1]{Fig.~\ref{#1}}

\usepackage{umoline}

\begin{document}

\title{Non-equilibrium dynamics of fluctuations in an ultra-cold atomic mixture}

\author{Apoorva Hegde }
\thanks{Both authors contributed equally to this work}
\affiliation{Heidelberg University, Kirchhoff Institute for Physics, Im Neuenheimer Feld 226, 69120 Heidelberg, Germany}
\author{Robert Ott}
\thanks{Both authors contributed equally to this work}
\affiliation{Heidelberg University, Institut f\"{u}r Theoretische Physik, Philosophenweg 16, 69120 Heidelberg, Germany}
\author{Andy Xia}
\affiliation{Heidelberg University, Kirchhoff Institute for Physics, Im Neuenheimer Feld 226, 69120 Heidelberg, Germany}
\author{Valentin Kasper}
\affiliation{ICFO-Institut de Ciencies Fotoniques, The Barcelona Institute of Science and Technology, Av. Carl Friedrich Gauss 3, 08860
Barcelona, Spain}
\author{Jürgen Berges}
\affiliation{Heidelberg University, Institut f\"{u}r Theoretische Physik, Philosophenweg 16, 69120 Heidelberg, Germany}
\author{Fred Jendrzejewski}
\affiliation{Heidelberg University, Kirchhoff Institute for Physics, Im Neuenheimer Feld 226, 69120 Heidelberg, Germany}
    
\begin{abstract}
We investigate an ultra-cold mixture of Bose gases interacting via spin-changing collisions by studying the dynamics of spin fluctuations. The experimental implementation employs $^{23}$Na and $^{7}$Li atoms, which are prepared out of equilibrium across a wide range of initial conditions. We identify three regimes in the dynamics of the system for different initial states: a long-lived metastable regime, an instability range with strong growth of fluctuations, and a fast relaxing regime approaching thermal equilibrium. Theoretical modelling of the data allows us to reconstruct effective potentials which characterize the different dynamical regimes of the system.
\end{abstract}
\maketitle

\textit{Introduction.} The experimental advances in ultra-cold atomic systems provide powerful platforms for the study of non-equilibrium dynamics of quantum many-body systems \cite{bernien2017probing, choi2016exploring,trotzky2012probing,gring2012relaxation,kaufman2016quantum,keesling2019quantum,lukin2019probing,schreiber2015observation,scherg2021observing}. While the mean-field dynamics frequently leads to a successful description of the many-body system, fluctuations can be crucial in out-of-equilibrium situations, such as instabilities~\cite{wintersperger2020parametric}, the build-up of structure far from equilibrium~\cite{prufer2018observation,erne2018universal,eigen2018universal,glidden2021bidirectional,navon2015critical,luschen2017observation}, or the fate of the ``false vacuum'' in tunneling processes~\cite{coleman1977fate}. Here, cold-atom simulators promise unique insights into the dynamics of fluctuations in quantum many-body systems~\cite{mouchet2001chaos,mehboudi2019using}.

\begin{figure}[h!]
	\includegraphics[width=0.92\columnwidth]{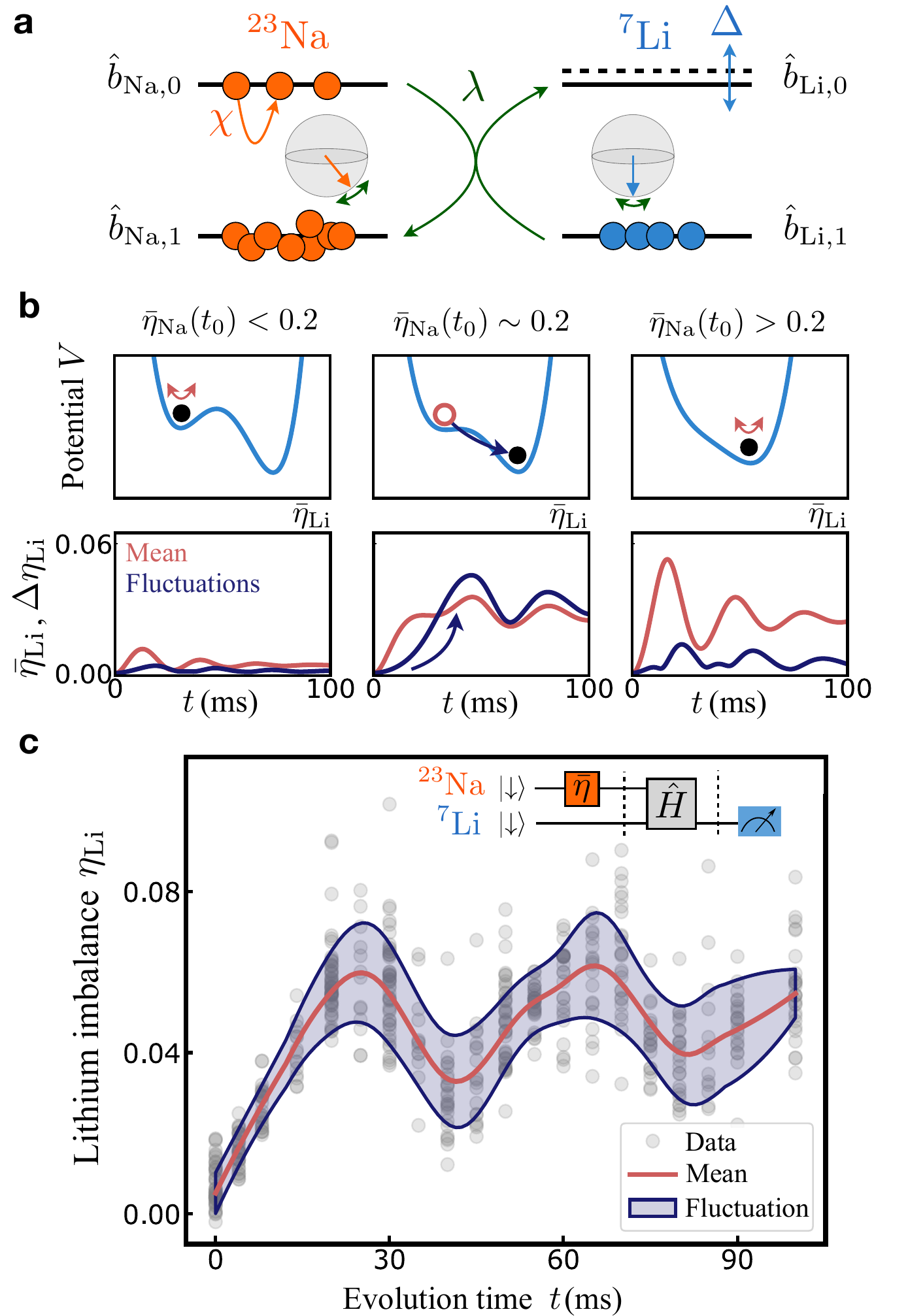}
	\caption{\textbf{Dynamics of spin fluctuations induced by spin changing collisions:} (\textbf{a}) The ultra-cold atomic mixture is coupled through spin-changing collisions between sodium and lithium atoms realizing effective spins degrees of freedom. (\textbf{b}) Varying initial conditions probe the mixture in three different regimes as illustrated with effective potentials: a  meta-stable state, an instability with growth of fluctuations, and fast relaxation. Solid lines show numerical real-time calculations of our effective model (SM). (\textbf{c})~The fluctuations of the data (shaded blue) around mean values (red) grow in time, where grey circles show single realizations of preparation, evolution and measurement (see inset). Solid lines are smoothed interpolations of data points.}
	\label{fig:Sketch}
\end{figure}

In this work, we investigate the non-equilibrium dynamics of an atomic mixture of  $^{23}$Na and $^{7}$Li atoms. Such cold atomic mixture experiments provide versatile platforms for simulations of different physical systems, ranging from the Bose polaron problem~\cite{hu2016bose,jorgensen2016observation,rentrop2016observation} to the quantum simulation of fundamental gauge theories~\cite{aidelsburger2022cold}. By using spin-changing collisions~\cite{Dajun2015} to couple the two atomic species, our setup implements a scalable building block of a lattice gauge theory~\cite{mil2020scalable,dai2017four,schweizer2019floquet}. While previously, the dynamics of the mean values were investigated with our setup~\cite{mil2020scalable}, here, we analyze the dynamics of spin fluctuations across a wide range of non-equilibrium initial conditions.

We identify different dynamical regimes of our system and characterize the evolution in terms of phenomena such as instabilities and the onset of thermalization. Depending on the initial spin imbalance of our mixture, we identify the three regimes: a long-lived metastable regime, an unstable region showing strong growth of fluctuations, and a fast relaxing regime approaching thermal equilibrium. We compare our data to theory by extending the earlier mean-field model~\cite{mil2020scalable} to incorporate the dynamics of fluctuations. Experimentally extracting the fluctuations allows us to reconstruct the effective potentials that characterize the different types of dynamical regimes. Our approach enables a detailed characterization of the complex many-body states out-of-equilibrium in atomic mixture experiments.

\textit{Experimental setup and state preparation.} The experiment is initiated by preparing an ultra-cold bosonic mixture of sodium (Na) and lithium (Li) atoms in a crossed optical dipole trap. After cooling and condensation, the sample consists of approximately $N_\mathrm{Na} \approx 3\times10^5$ sodium and $N_\mathrm{Li} \approx 3\times10^4$ lithium atoms in overlapping atomic clouds. The two species are prepared in their spin $F=1$ manifolds, where the application of an external magnetic field ($B= 2.118(2)$G) leads to an energy splitting between three internal hyperfine states \cite{stamper2013spinor}. Suitably choosing such energy separations allows us to confine the atoms to their lowest lying states, $\ket{F=1,m_F=0,1}$, for both species, see \Fig{fig:Sketch}\textbf{a}.

Initially, the atoms are prepared in $m_F=1$ corresponding to their single-particle ground state. To initialize the non-equilibrium dynamics, we follow the experimental sequence described in Ref.~\cite{mil2020scalable}. We transfer an average fraction $\bar{\eta}_\mathrm{Na}$ of Na atoms to the energetically higher $m_F = 0$ state, thereby exciting the atoms out of equilibrium. To achieve this, the internal Na states are coupled with a highly tunable microwave setup, where a two-pulse sequence transfers the atoms through the intermediate state $\ket{F=2,m_F=2}$. While this provides an efficient passage for Na atoms, corresponding internal states of Li are kept out of resonance, and it remains polarized in $\ket{F=1,m_F=1}$ for all considered initial states at time $t_0=0$ms. We prepare a wide range of Na initial states $\bar{\eta}_\mathrm{Na}(t_0)$, which also sets the total (conserved) magnetization for the dynamics, which is the sum of the population difference of the respective magnetic substates. The different initial states allow us to tune the effective system parameters to explore non-equilibrium physics in different regimes. The various types of non-equilibrium phenomena are illustrated in \Fig{fig:Sketch}\textbf{b}, where the dynamics can be understood from effective potentials.

\textit{Non-equilibrium dynamics.} After state preparation, we let the system evolve for evolution times of up to $t=100$ms, during which the atoms exchange spin, resulting in oscillatory dynamics as shown in Fig.~\ref{fig:Sketch}\textbf{c}. The spin dynamics is made possible through a precise control of the external magnetic field, such that spin states of Na and Li are kept in resonance during the evolution. To investigate the complex dynamics, we detect the resulting spin populations of lithium by performing state-selective time-of-flight (TOF) measurements (``Stern-Gerlach method"). Here, atoms in different spin states are spatially separated with a magnetic field gradient before imaging, which gives access to the individual atom populations $N_{\mathrm{Li},m_F}$ and hence the spin imbalance $\hat{\eta}_\mathrm{Li} = \hat{N}_{\mathrm{Li},0}/\hat{N}_\mathrm{Li}$, where $\hat{N}_\mathrm{Li}=\hat{N}_{\mathrm{Li},0}+\hat{N}_{\mathrm{Li},1}$, as shown in Fig.~\ref{fig:Sketch}\textbf{b}. We repeat the experimental cycle up to 20 times per evolution time $t$ to observe the fluctuating signal. From our measurements, we extract the mean values $\bar{\eta}_\mathrm{Li} = \langle \hat{\eta}_\mathrm{Li}\rangle$ and second-order correlations $\Delta\eta_\mathrm{Li}= \sqrt{\langle \hat{\eta}_\mathrm{Li}\hat{\eta}_\mathrm{Li}\rangle -\langle \hat{\eta}_\mathrm{Li}\rangle ^2}$ in order to characterize the non-equilibrium dynamics~\footnote{Definitions of imbalances for Na are analogous.}.

\begin{figure}[t!]
	\includegraphics[width=0.45\textwidth]{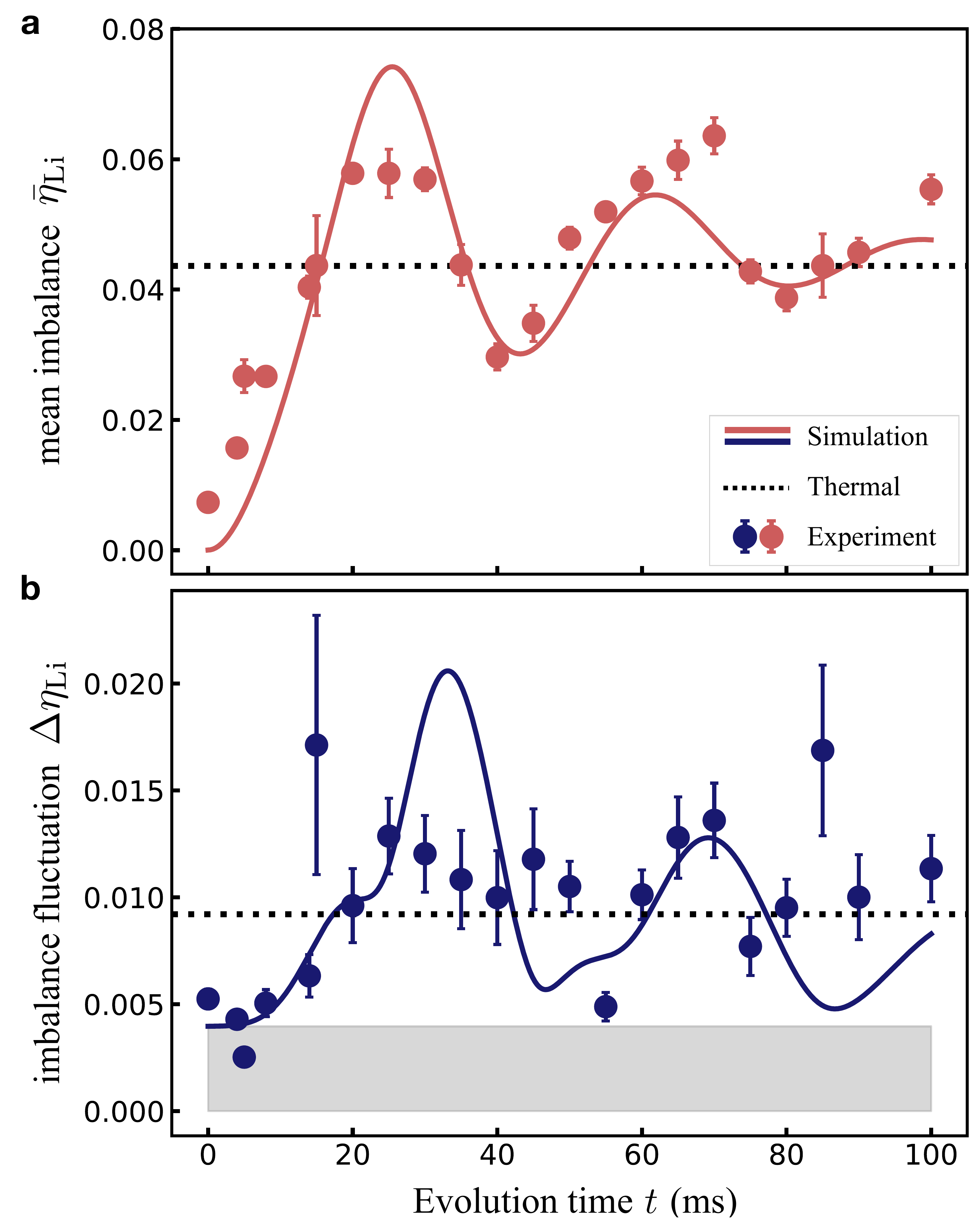}
	\caption{\textbf{Time evolution of mean and fluctuations:} (\textbf{a})~We show the evolution of mean lithium spin transfer $\bar{\eta}_\mathrm{L}$ for evolution times $t\leq 100$ms starting from a mean sodium imbalance of $\bar{\eta}_\mathrm{Na} =0.406$. The damped-oscillatory behaviour of the data (red circles) is well-described by our semi-classical calculation (solid line) which accommodates initial state fluctuations. (\textbf{b})~The two-point correlation data (blue circles) shows an initial growth of fluctuations (see \Fig{fig:Sketch}\textbf{c}) and reaches an approximately constant value at later times. The late time data is in approximate agreement with a thermal calculation for a temperature $T=630$nK (dotted line). Error bars represent standard error on the mean. Our detection of spin fluctuations has a systematic offset of $0.0040(1)$ for~$\Delta\bar{\eta}_\mathrm{Li}$ (grey shaded area), which we added to the numerical curve in \textbf{b}.}
	\label{fig:Dynamics}
\end{figure}

For the initial state with $\bar{\eta}_\mathrm{Na}=0.4$, the experimentally observed spin evolution is displayed in Fig.~\ref{fig:Dynamics}. As panel \textbf{a} shows, the signal of the mean Li imbalance $\bar{\eta}_\mathrm{Li}$ rises rapidly initially, and it is strongly damped towards a steady state at late times. Furthermore, we investigate the data beyond mean populations by considering the dynamics of corresponding imbalance fluctuations $\Delta\bar{\eta}_\mathrm{Li}$, see panel \textbf{b}. Fluctuations, being small initially, build up quickly during the first $30$ms, and they approach an approximately constant value of $\Delta\eta_\mathrm{Li} \approx 0.01$ at later times.

\textit{Effective spin model.} To diagnose relevant processes and access the effective potential landscape of the many-body problem, we compare our measurements to numerical real-time calculations. In the experiment, the atoms are localized in a deep harmonic trap. Hence, we describe the coupled system of sodium ($a=\mathrm{Na}$) and lithium ($a=\mathrm{Li}$) with single spatial modes for each magnetic substate $m_F = 0,1$, where the bosonic creation (annihilation) operators $\hat{b}^\dagger_{a,i}$ ($\hat{b}_{a,i}$)
fulfill commutation relations $[\hat{b}_{a,i},\hat{b}_{b,j}^\dagger] = \delta_{ij}\delta_{ab}$. In view of the atoms' SU(2) symmetry, we introduce collective spin operators using the Schwinger boson representation, i.e. $\hat{L}_{a,+}=\hat{b}^\dagger_{a,0}\hat{b}_{a,1}$, $\hat{L}_{a,-}=\hat{b}_{a,0}\hat{b}_{a,1}^\dagger$, and $\hat{L}_{a,z}=(\hat{b}^\dagger_{a,0}\hat{b}_{a,0} -\hat{b}^\dagger_{a,1}\hat{b}_{a,1})/2$. The evolution is governed by the Hamiltonian (see Supplemental Material (SM) for details)
\begin{align} \label{eq:Hamiltonian}
    \hat{H}=\chi \hat{L}_{\mathrm{Na},z}^{2}+\Delta \hat{L}_{\mathrm{Li},z}+\lambda\big(\hat{L}_{\mathrm{Na},+}\hat{L}_{\mathrm{Li},-}+ h.c.\big),
\end{align}
where $h.c.$ denotes hermitian conjugation. The spin-changing dynamics is driven by the interaction coupling $\lambda$. The term $\Delta$ models the effective energy offset between magnetic substates, and $\chi$ characterizes intra and inter-species interaction strengths from atom collisions without changes in spin. Here, $\Delta$ depends only on the total numbers of Na and Li atoms and the total magnetization $\hat{M} = \hat{L}_{\mathrm{Li},z}+\hat{L}_{\mathrm{Na},z}$, which are modelled as conserved quantities during the evolution, but depend on the details of the initial condition. Carefully choosing these conserved quantities hence enables the controlled investigation of our system in different parameter regimes. We consider a semi-classical (truncated Wigner) approximation to describe the large condensates realized in the experiment. The effective spin degrees of freedom are simulated with classical evolution equations, and quantum effects are implemented as the Gaussian initial state fluctuations of the initial coherent spin states. Furthermore, we average over initial atom number fluctuations, which vary for different realizations by about $10\%$ for both Na and Li. The initial state fluctuations and the model parameters are detailed in the SM.

In addition to the coherent spin dynamics, the spatial excitations of the atomic cloud are expected to interact with spin degrees of freedom and exchange energy. To account for such an effect in our model, we include a friction term for the spin as given by a Landau-Lifshitz-Gilbert-type damping~\cite{garcia1998langevin} $\partial_t \mathbf{L}_{\mathrm{Li}} = \{\mathbf{L}_{\mathrm{Li}},H\} -\gamma( L_{\mathrm{Li},z}\mathbf{L}_\mathrm{Li} - \mathbf{L}_\mathrm{Li}^2\mathbf{e}_{z} )$, where $\mathbf{L}_{\mathrm{Li}}$ is the three-component lithium spin, $\mathbf{e}_{z}$ is the unit vector in $z$ direction, and $\{\mathbf{L}_{\mathrm{Li}},H\}$ represents the (undamped) classical evolution equations via Poisson brackets. Here, we choose a damping rate of $\gamma = 1.8\times 10^{-3}$Hz, and furthermore adjust the sodium spin accordingly to conserve the system's total magnetization, see SM. With this friction term, we observe a fast damping at long times, see~\Fig{fig:Dynamics}.

 To characterize the dynamics observed in the present experiment, we compare our data with thermal equilibrium. To this end, we define a canonical ensemble by the partition sum $Z = \mathrm{Tr} [ \hat{P}_M \exp(-\beta\hat{H})]$, where $\hat{P}_M$ projects the system to a fixed magnetization $\bra{\psi_0}\hat{M}\ket{\psi_0}$, which is set by the initial state $\ket{\psi_0}$. Furthermore, $\beta = 1/(k_B T)$ sets the (inverse) temperature of the system, where $k_B$ is the Boltzmann constant. In \Fig{fig:Dynamics} we compare our results to a thermal ensemble with the temperature of the condensate $T = \SI{630(20)}{\nano \kelvin}$ as obtained by extracting the atoms' momentum distributions with TOF-measurements. At late times, both mean value and fluctuations of the Li imbalance oscillate around these thermal estimates.

\begin{figure}[t!]
	\includegraphics[width=0.45\textwidth]{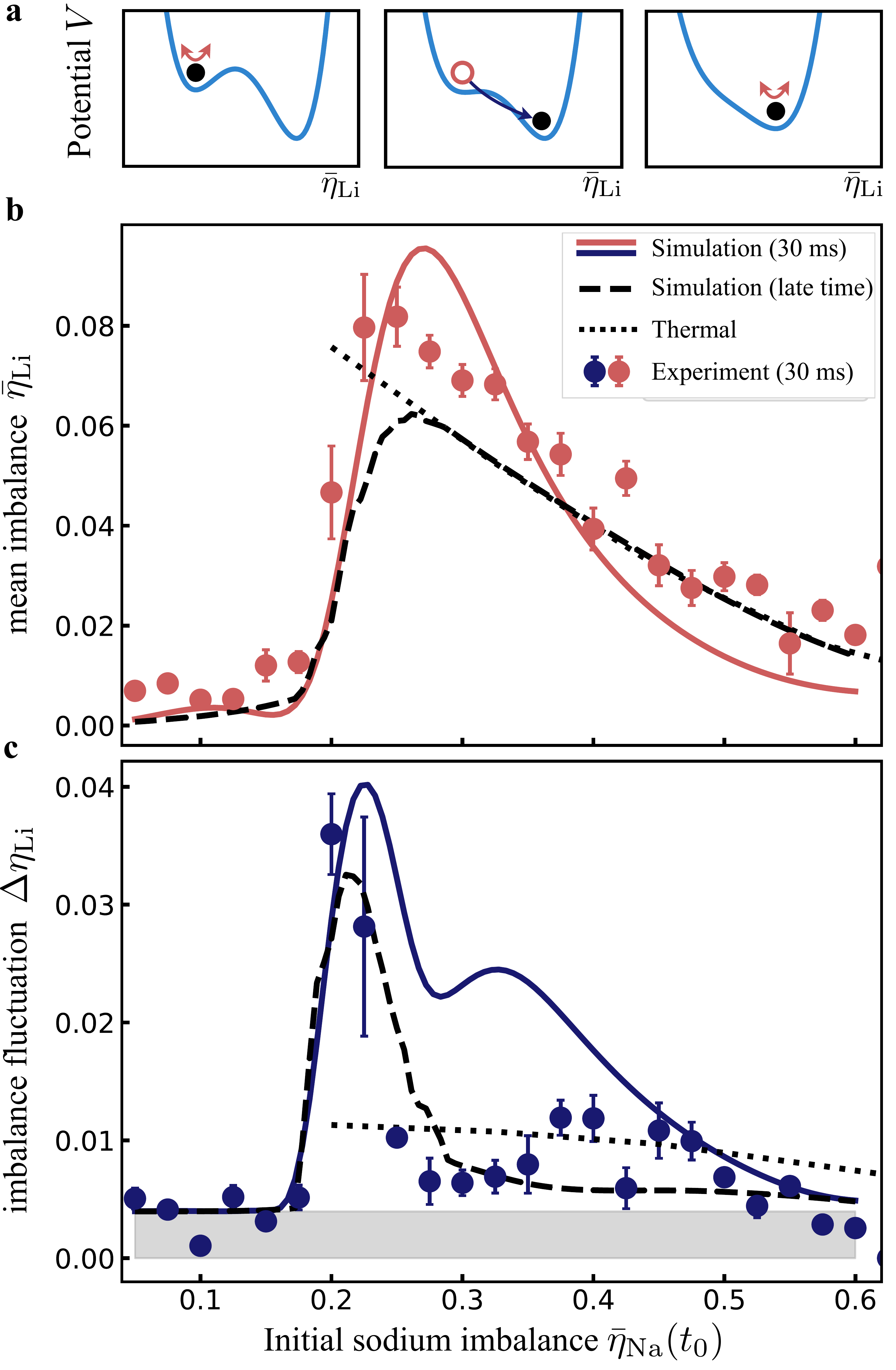}
	\caption{\textbf{Non-equilibrium regimes:} (\textbf{a})~We indicate the different dynamical regimes with sketches of the potential landscape (see main text). (\textbf{b})~After $30$ms evolution time, the mean lithium imbalance $\bar{\eta}_\mathrm{Li}$ reveals a sharp resonance behavior for changing initial conditions set by $\bar{\eta}_\mathrm{Na}$. (\textbf{c}) The fluctuation signal shows a large peak around $\bar{\eta}_\mathrm{Li} \sim 0.2$, indicating an instability regime characterized by the fast growth of fluctuations. We find an overall good agreement with numerical real-time calculations at $30$ms (solid colored) and a late time calculation at $150$ ms (dashed). On the right-hand side of the transition we find good agreement with a (classical) thermal ensemble at temperature $630(20)$nK (solid black). The shaded area in (\textbf{c}) indicates the fluctuation offset from detection noise, which we add to the numerical curve in \textbf{c}.}
	\label{fig:Imbalance}
\end{figure}

\textit{Probing the non-equilibrium regimes.} To explore the dynamics in different parameter regimes, we consider the system evolution for different initial states in the following.
In \Fig{fig:Imbalance} we show the Li imbalance after $30$ms evolution time for a range of Na initial states. We observe a pronounced resonance shape in the mean imbalance value, with a sharp transition around $ \bar{\eta}_\mathrm{Na} \approx 0.2$, as illustrated in panel \textbf{b}. On the left side of the transition, $\bar{\eta}_\mathrm{Na} < 0.2$, the Li imbalance stays small. On the other hand, it decays on the right side of the resonance for larger values of $\bar{\eta}_\mathrm{Na}$. The transition is accompanied with a sharp peak in the fluctuations $\Delta \eta_\mathrm{Li}$, as displayed in panel \textbf{c}. This peak indicates an enhanced sensitivity of the dynamics, leading to a fast growth of fluctuations in the data.
The data is in good agreement with the results from our semi-classical numerical calculations, as shown in \Fig{fig:Imbalance}, and our calculations recover the essential features of the spin-changing dynamics.

As illustrated for $\bar{\eta}_\mathrm{Na} = 0.4$ in \Fig{fig:Dynamics}, data taken as early as $30$ms may already provide valuable insight for the approximate late-time evolution of the system. In fact, for the given initial conditions, the experimental data in \Fig{fig:Imbalance} shows an excellent agreement with late time numerical results at t=150ms. To characterize our system in the regime $\bar{\eta}_\mathrm{Na} > 0.2$, we compare the data to corresponding predictions from the thermal ensemble, finding an overall good agreement on the right-hand side of the resonance shape, where the agreement is especially precise on the mean imbalance of Li. We thus observe indications for an onset of thermalization in our system. Conversely, on the left-hand side of the transition the spin-changing dynamics is strongly suppressed, and thermalization dynamics is not initiated on observable time scales. In our theoretical modelling, this suppression extends to evolution times beyond 150ms, indicating a long-lived metastable state.

These qualitative features of the system dynamics may be illustrated in terms of an evolution within effective potentials. For fixed conserved charges, the classical evolution of the Li imbalance, as relevant for the early time dynamics, may be written as a second-order differential equation $\partial_t^2 \bar{\eta}_\mathrm{Li} = -\partial V(\bar{\eta}_\mathrm{Li})/\partial\bar{\eta}_\mathrm{Li}$ with a quartic potential $V$~\cite{torsten2020}, see SM. In \Fig{fig:Imbalance} we show three configurations of the potential for different values of the total magnetization as set by the initial state of Na.
For small initial Na imbalances $\bar{\eta}_\mathrm{Na} < 0.2$ the system is constrained to small Li imbalances, as indicated by a second, local minimum in the effective potential (left). Conversely, for larger $\bar{\eta}_\mathrm{Na}> 0.2$, the system evolution is governed by oscillations around a single global minimum which approximately corresponds to the thermal mean value.
These regimes are separated by a sharp transition at $\bar{\eta}_\mathrm{Na}\sim 0.2$, where the system dynamics becomes unstable with respect to the second minimum. Corresponding mean and fluctuation values are expected to rapidly rise to similar magnitudes, as confirmed by the data in \Fig{fig:Imbalance}. 

\textit{Conclusion.} In summary, we investigated the real-time dynamics of an ultra-cold Bose mixture out of equilibrium. This was enabled through the experimental extraction of fluctuations, which provides a powerful tool for the characterization of many-body quantum systems in \cite{schweigler2017experimental,zache2020extracting} and out of equilibrium~\cite{rispoli2019quantum,prufer2020experimental}. An exciting prospect for future work is the controlled investigation of correlation functions in cold-atom gauge field theories. Our present system realizes a minimal building block of a U(1) gauge theory, which may be extended in one- or two spatial dimensions with optical lattices~\cite{mil2020scalable,ott2021scalable}. This will also open up new avenues towards quantum simulations of the complex real-time dynamics of gauge theories~\cite{surace2020lattice,gorg2019realization,martinez2016real,schweizer2019floquet,klco2018quantum,zhou2021thermalization,mildenberger2022probing,bernien2017probing}.

\textit{Acknowledgments.} We thank A. Garcia Sala and T. V. Zache for fruitful discussions and collaboration on related work. This work is supported by the DFG
Collaborative Research Centre “SFB 1225 (ISOQUANT)" under Project-ID 27381115. F.J. acknowledges the DFG support through the
project FOR 2724 and the Emmy-Noether grant (Project-
ID 377616843).

ICFO group acknowledges support from: ERC AdG NOQIA; Agencia Estatal de Investigación (R\&D project CEX2019-000910-S, funded by MCIN/ AEI/10.13039/501100011033, Plan National FIDEUA PID2019-106901GB-I00, FPI, QUANTERA MAQS PCI2019-111828-2, Proyectos de I+D+I “Retos Colaboración” QUSPIN RTC2019-007196-7);  Fundació Cellex; Fundació Mir-Puig; Generalitat de Catalunya through the European Social Fund FEDER and CERCA program (AGAUR Grant No. 2017 SGR 134, QuantumCAT \ U16-011424, co-funded by ERDF Operational Program of Catalonia 2014-2020); EU Horizon 2020 FET-OPEN OPTOlogic (Grant No 899794); National Science Centre, Poland (Symfonia Grant No. 2016/20/W/ST4/00314); European Union’s Horizon 2020 research and innovation programme under the Marie-Skłodowska-Curie grant agreement No 101029393 (STREDCH) and No 847648  (“La Caixa” Junior Leaders fellowships ID100010434: LCF/BQ/PI19/11690013, LCF/BQ/PI20/11760031,  LCF/BQ/PR20/11770012, LCF/BQ/PR21/11840013). 

\bibliographystyle{apsrev4-1} 
\bibliography{bibliography}

\begin{thebibliography}{42}%
\makeatletter
\providecommand \@ifxundefined [1]{%
 \@ifx{#1\undefined}
}%
\providecommand \@ifnum [1]{%
 \ifnum #1\expandafter \@firstoftwo
 \else \expandafter \@secondoftwo
 \fi
}%
\providecommand \@ifx [1]{%
 \ifx #1\expandafter \@firstoftwo
 \else \expandafter \@secondoftwo
 \fi
}%
\providecommand \natexlab [1]{#1}%
\providecommand \enquote  [1]{``#1''}%
\providecommand \bibnamefont  [1]{#1}%
\providecommand \bibfnamefont [1]{#1}%
\providecommand \citenamefont [1]{#1}%
\providecommand \href@noop [0]{\@secondoftwo}%
\providecommand \href [0]{\begingroup \@sanitize@url \@href}%
\providecommand \@href[1]{\@@startlink{#1}\@@href}%
\providecommand \@@href[1]{\endgroup#1\@@endlink}%
\providecommand \@sanitize@url [0]{\catcode `\\12\catcode `\$12\catcode
  `\&12\catcode `\#12\catcode `\^12\catcode `\_12\catcode `\%12\relax}%
\providecommand \@@startlink[1]{}%
\providecommand \@@endlink[0]{}%
\providecommand \url  [0]{\begingroup\@sanitize@url \@url }%
\providecommand \@url [1]{\endgroup\@href {#1}{\urlprefix }}%
\providecommand \urlprefix  [0]{URL }%
\providecommand \Eprint [0]{\href }%
\providecommand \doibase [0]{http://dx.doi.org/}%
\providecommand \selectlanguage [0]{\@gobble}%
\providecommand \bibinfo  [0]{\@secondoftwo}%
\providecommand \bibfield  [0]{\@secondoftwo}%
\providecommand \translation [1]{[#1]}%
\providecommand \BibitemOpen [0]{}%
\providecommand \bibitemStop [0]{}%
\providecommand \bibitemNoStop [0]{.\EOS\space}%
\providecommand \EOS [0]{\spacefactor3000\relax}%
\providecommand \BibitemShut  [1]{\csname bibitem#1\endcsname}%
\let\auto@bib@innerbib\@empty
\bibitem [{\citenamefont {Bernien}\ \emph {et~al.}(2017)\citenamefont
  {Bernien}, \citenamefont {Schwartz}, \citenamefont {Keesling}, \citenamefont
  {Levine}, \citenamefont {Omran}, \citenamefont {Pichler}, \citenamefont
  {Choi}, \citenamefont {Zibrov}, \citenamefont {Endres}, \citenamefont
  {Greiner} \emph {et~al.}}]{bernien2017probing}%
  \BibitemOpen
  \bibfield  {author} {\bibinfo {author} {\bibfnamefont {H.}~\bibnamefont
  {Bernien}}, \bibinfo {author} {\bibfnamefont {S.}~\bibnamefont {Schwartz}},
  \bibinfo {author} {\bibfnamefont {A.}~\bibnamefont {Keesling}}, \bibinfo
  {author} {\bibfnamefont {H.}~\bibnamefont {Levine}}, \bibinfo {author}
  {\bibfnamefont {A.}~\bibnamefont {Omran}}, \bibinfo {author} {\bibfnamefont
  {H.}~\bibnamefont {Pichler}}, \bibinfo {author} {\bibfnamefont
  {S.}~\bibnamefont {Choi}}, \bibinfo {author} {\bibfnamefont {A.~S.}\
  \bibnamefont {Zibrov}}, \bibinfo {author} {\bibfnamefont {M.}~\bibnamefont
  {Endres}}, \bibinfo {author} {\bibfnamefont {M.}~\bibnamefont {Greiner}},
  \emph {et~al.},\ }\href@noop {} {\bibfield  {journal} {\bibinfo  {journal}
  {Nature}\ }\textbf {\bibinfo {volume} {551}},\ \bibinfo {pages} {579}
  (\bibinfo {year} {2017})}\BibitemShut {NoStop}%
\bibitem [{\citenamefont {Choi}\ \emph {et~al.}(2016)\citenamefont {Choi},
  \citenamefont {Hild}, \citenamefont {Zeiher}, \citenamefont {Schau{\ss}},
  \citenamefont {Rubio-Abadal}, \citenamefont {Yefsah}, \citenamefont
  {Khemani}, \citenamefont {Huse}, \citenamefont {Bloch},\ and\ \citenamefont
  {Gross}}]{choi2016exploring}%
  \BibitemOpen
  \bibfield  {author} {\bibinfo {author} {\bibfnamefont {J.-y.}\ \bibnamefont
  {Choi}}, \bibinfo {author} {\bibfnamefont {S.}~\bibnamefont {Hild}}, \bibinfo
  {author} {\bibfnamefont {J.}~\bibnamefont {Zeiher}}, \bibinfo {author}
  {\bibfnamefont {P.}~\bibnamefont {Schau{\ss}}}, \bibinfo {author}
  {\bibfnamefont {A.}~\bibnamefont {Rubio-Abadal}}, \bibinfo {author}
  {\bibfnamefont {T.}~\bibnamefont {Yefsah}}, \bibinfo {author} {\bibfnamefont
  {V.}~\bibnamefont {Khemani}}, \bibinfo {author} {\bibfnamefont {D.~A.}\
  \bibnamefont {Huse}}, \bibinfo {author} {\bibfnamefont {I.}~\bibnamefont
  {Bloch}}, \ and\ \bibinfo {author} {\bibfnamefont {C.}~\bibnamefont
  {Gross}},\ }\href@noop {} {\bibfield  {journal} {\bibinfo  {journal}
  {Science}\ }\textbf {\bibinfo {volume} {352}},\ \bibinfo {pages} {1547}
  (\bibinfo {year} {2016})}\BibitemShut {NoStop}%
\bibitem [{\citenamefont {Trotzky}\ \emph {et~al.}(2012)\citenamefont
  {Trotzky}, \citenamefont {Chen}, \citenamefont {Flesch}, \citenamefont
  {McCulloch}, \citenamefont {Schollw{\"o}ck}, \citenamefont {Eisert},\ and\
  \citenamefont {Bloch}}]{trotzky2012probing}%
  \BibitemOpen
  \bibfield  {author} {\bibinfo {author} {\bibfnamefont {S.}~\bibnamefont
  {Trotzky}}, \bibinfo {author} {\bibfnamefont {Y.-A.}\ \bibnamefont {Chen}},
  \bibinfo {author} {\bibfnamefont {A.}~\bibnamefont {Flesch}}, \bibinfo
  {author} {\bibfnamefont {I.~P.}\ \bibnamefont {McCulloch}}, \bibinfo {author}
  {\bibfnamefont {U.}~\bibnamefont {Schollw{\"o}ck}}, \bibinfo {author}
  {\bibfnamefont {J.}~\bibnamefont {Eisert}}, \ and\ \bibinfo {author}
  {\bibfnamefont {I.}~\bibnamefont {Bloch}},\ }\href@noop {} {\bibfield
  {journal} {\bibinfo  {journal} {Nature physics}\ }\textbf {\bibinfo {volume}
  {8}},\ \bibinfo {pages} {325} (\bibinfo {year} {2012})}\BibitemShut {NoStop}%
\bibitem [{\citenamefont {Gring}\ \emph {et~al.}(2012)\citenamefont {Gring},
  \citenamefont {Kuhnert}, \citenamefont {Langen}, \citenamefont {Kitagawa},
  \citenamefont {Rauer}, \citenamefont {Schreitl}, \citenamefont {Mazets},
  \citenamefont {Smith}, \citenamefont {Demler},\ and\ \citenamefont
  {Schmiedmayer}}]{gring2012relaxation}%
  \BibitemOpen
  \bibfield  {author} {\bibinfo {author} {\bibfnamefont {M.}~\bibnamefont
  {Gring}}, \bibinfo {author} {\bibfnamefont {M.}~\bibnamefont {Kuhnert}},
  \bibinfo {author} {\bibfnamefont {T.}~\bibnamefont {Langen}}, \bibinfo
  {author} {\bibfnamefont {T.}~\bibnamefont {Kitagawa}}, \bibinfo {author}
  {\bibfnamefont {B.}~\bibnamefont {Rauer}}, \bibinfo {author} {\bibfnamefont
  {M.}~\bibnamefont {Schreitl}}, \bibinfo {author} {\bibfnamefont
  {I.}~\bibnamefont {Mazets}}, \bibinfo {author} {\bibfnamefont {D.~A.}\
  \bibnamefont {Smith}}, \bibinfo {author} {\bibfnamefont {E.}~\bibnamefont
  {Demler}}, \ and\ \bibinfo {author} {\bibfnamefont {J.}~\bibnamefont
  {Schmiedmayer}},\ }\href@noop {} {\bibfield  {journal} {\bibinfo  {journal}
  {Science}\ }\textbf {\bibinfo {volume} {337}},\ \bibinfo {pages} {1318}
  (\bibinfo {year} {2012})}\BibitemShut {NoStop}%
\bibitem [{\citenamefont {Kaufman}\ \emph {et~al.}(2016)\citenamefont
  {Kaufman}, \citenamefont {Tai}, \citenamefont {Lukin}, \citenamefont
  {Rispoli}, \citenamefont {Schittko}, \citenamefont {Preiss},\ and\
  \citenamefont {Greiner}}]{kaufman2016quantum}%
  \BibitemOpen
  \bibfield  {author} {\bibinfo {author} {\bibfnamefont {A.~M.}\ \bibnamefont
  {Kaufman}}, \bibinfo {author} {\bibfnamefont {M.~E.}\ \bibnamefont {Tai}},
  \bibinfo {author} {\bibfnamefont {A.}~\bibnamefont {Lukin}}, \bibinfo
  {author} {\bibfnamefont {M.}~\bibnamefont {Rispoli}}, \bibinfo {author}
  {\bibfnamefont {R.}~\bibnamefont {Schittko}}, \bibinfo {author}
  {\bibfnamefont {P.~M.}\ \bibnamefont {Preiss}}, \ and\ \bibinfo {author}
  {\bibfnamefont {M.}~\bibnamefont {Greiner}},\ }\href@noop {} {\bibfield
  {journal} {\bibinfo  {journal} {Science}\ }\textbf {\bibinfo {volume}
  {353}},\ \bibinfo {pages} {794} (\bibinfo {year} {2016})}\BibitemShut
  {NoStop}%
\bibitem [{\citenamefont {Keesling}\ \emph {et~al.}(2019)\citenamefont
  {Keesling}, \citenamefont {Omran}, \citenamefont {Levine}, \citenamefont
  {Bernien}, \citenamefont {Pichler}, \citenamefont {Choi}, \citenamefont
  {Samajdar}, \citenamefont {Schwartz}, \citenamefont {Silvi}, \citenamefont
  {Sachdev} \emph {et~al.}}]{keesling2019quantum}%
  \BibitemOpen
  \bibfield  {author} {\bibinfo {author} {\bibfnamefont {A.}~\bibnamefont
  {Keesling}}, \bibinfo {author} {\bibfnamefont {A.}~\bibnamefont {Omran}},
  \bibinfo {author} {\bibfnamefont {H.}~\bibnamefont {Levine}}, \bibinfo
  {author} {\bibfnamefont {H.}~\bibnamefont {Bernien}}, \bibinfo {author}
  {\bibfnamefont {H.}~\bibnamefont {Pichler}}, \bibinfo {author} {\bibfnamefont
  {S.}~\bibnamefont {Choi}}, \bibinfo {author} {\bibfnamefont {R.}~\bibnamefont
  {Samajdar}}, \bibinfo {author} {\bibfnamefont {S.}~\bibnamefont {Schwartz}},
  \bibinfo {author} {\bibfnamefont {P.}~\bibnamefont {Silvi}}, \bibinfo
  {author} {\bibfnamefont {S.}~\bibnamefont {Sachdev}},  \emph {et~al.},\
  }\href@noop {} {\bibfield  {journal} {\bibinfo  {journal} {Nature}\ }\textbf
  {\bibinfo {volume} {568}},\ \bibinfo {pages} {207} (\bibinfo {year}
  {2019})}\BibitemShut {NoStop}%
\bibitem [{\citenamefont {Lukin}\ \emph {et~al.}(2019)\citenamefont {Lukin},
  \citenamefont {Rispoli}, \citenamefont {Schittko}, \citenamefont {Tai},
  \citenamefont {Kaufman}, \citenamefont {Choi}, \citenamefont {Khemani},
  \citenamefont {L{\'e}onard},\ and\ \citenamefont
  {Greiner}}]{lukin2019probing}%
  \BibitemOpen
  \bibfield  {author} {\bibinfo {author} {\bibfnamefont {A.}~\bibnamefont
  {Lukin}}, \bibinfo {author} {\bibfnamefont {M.}~\bibnamefont {Rispoli}},
  \bibinfo {author} {\bibfnamefont {R.}~\bibnamefont {Schittko}}, \bibinfo
  {author} {\bibfnamefont {M.~E.}\ \bibnamefont {Tai}}, \bibinfo {author}
  {\bibfnamefont {A.~M.}\ \bibnamefont {Kaufman}}, \bibinfo {author}
  {\bibfnamefont {S.}~\bibnamefont {Choi}}, \bibinfo {author} {\bibfnamefont
  {V.}~\bibnamefont {Khemani}}, \bibinfo {author} {\bibfnamefont
  {J.}~\bibnamefont {L{\'e}onard}}, \ and\ \bibinfo {author} {\bibfnamefont
  {M.}~\bibnamefont {Greiner}},\ }\href@noop {} {\bibfield  {journal} {\bibinfo
   {journal} {Science}\ }\textbf {\bibinfo {volume} {364}},\ \bibinfo {pages}
  {256} (\bibinfo {year} {2019})}\BibitemShut {NoStop}%
\bibitem [{\citenamefont {Schreiber}\ \emph {et~al.}(2015)\citenamefont
  {Schreiber}, \citenamefont {Hodgman}, \citenamefont {Bordia}, \citenamefont
  {L{\"u}schen}, \citenamefont {Fischer}, \citenamefont {Vosk}, \citenamefont
  {Altman}, \citenamefont {Schneider},\ and\ \citenamefont
  {Bloch}}]{schreiber2015observation}%
  \BibitemOpen
  \bibfield  {author} {\bibinfo {author} {\bibfnamefont {M.}~\bibnamefont
  {Schreiber}}, \bibinfo {author} {\bibfnamefont {S.~S.}\ \bibnamefont
  {Hodgman}}, \bibinfo {author} {\bibfnamefont {P.}~\bibnamefont {Bordia}},
  \bibinfo {author} {\bibfnamefont {H.~P.}\ \bibnamefont {L{\"u}schen}},
  \bibinfo {author} {\bibfnamefont {M.~H.}\ \bibnamefont {Fischer}}, \bibinfo
  {author} {\bibfnamefont {R.}~\bibnamefont {Vosk}}, \bibinfo {author}
  {\bibfnamefont {E.}~\bibnamefont {Altman}}, \bibinfo {author} {\bibfnamefont
  {U.}~\bibnamefont {Schneider}}, \ and\ \bibinfo {author} {\bibfnamefont
  {I.}~\bibnamefont {Bloch}},\ }\href@noop {} {\bibfield  {journal} {\bibinfo
  {journal} {Science}\ }\textbf {\bibinfo {volume} {349}},\ \bibinfo {pages}
  {842} (\bibinfo {year} {2015})}\BibitemShut {NoStop}%
\bibitem [{\citenamefont {Scherg}\ \emph {et~al.}(2021)\citenamefont {Scherg},
  \citenamefont {Kohlert}, \citenamefont {Sala}, \citenamefont {Pollmann},
  \citenamefont {Hebbe~Madhusudhana}, \citenamefont {Bloch},\ and\
  \citenamefont {Aidelsburger}}]{scherg2021observing}%
  \BibitemOpen
  \bibfield  {author} {\bibinfo {author} {\bibfnamefont {S.}~\bibnamefont
  {Scherg}}, \bibinfo {author} {\bibfnamefont {T.}~\bibnamefont {Kohlert}},
  \bibinfo {author} {\bibfnamefont {P.}~\bibnamefont {Sala}}, \bibinfo {author}
  {\bibfnamefont {F.}~\bibnamefont {Pollmann}}, \bibinfo {author}
  {\bibfnamefont {B.}~\bibnamefont {Hebbe~Madhusudhana}}, \bibinfo {author}
  {\bibfnamefont {I.}~\bibnamefont {Bloch}}, \ and\ \bibinfo {author}
  {\bibfnamefont {M.}~\bibnamefont {Aidelsburger}},\ }\href@noop {} {\bibfield
  {journal} {\bibinfo  {journal} {Nature Communications}\ }\textbf {\bibinfo
  {volume} {12}},\ \bibinfo {pages} {1} (\bibinfo {year} {2021})}\BibitemShut
  {NoStop}%
\bibitem [{\citenamefont {Wintersperger}\ \emph {et~al.}(2020)\citenamefont
  {Wintersperger}, \citenamefont {Bukov}, \citenamefont {N{\"a}ger},
  \citenamefont {Lellouch}, \citenamefont {Demler}, \citenamefont {Schneider},
  \citenamefont {Bloch}, \citenamefont {Goldman},\ and\ \citenamefont
  {Aidelsburger}}]{wintersperger2020parametric}%
  \BibitemOpen
  \bibfield  {author} {\bibinfo {author} {\bibfnamefont {K.}~\bibnamefont
  {Wintersperger}}, \bibinfo {author} {\bibfnamefont {M.}~\bibnamefont
  {Bukov}}, \bibinfo {author} {\bibfnamefont {J.}~\bibnamefont {N{\"a}ger}},
  \bibinfo {author} {\bibfnamefont {S.}~\bibnamefont {Lellouch}}, \bibinfo
  {author} {\bibfnamefont {E.}~\bibnamefont {Demler}}, \bibinfo {author}
  {\bibfnamefont {U.}~\bibnamefont {Schneider}}, \bibinfo {author}
  {\bibfnamefont {I.}~\bibnamefont {Bloch}}, \bibinfo {author} {\bibfnamefont
  {N.}~\bibnamefont {Goldman}}, \ and\ \bibinfo {author} {\bibfnamefont
  {M.}~\bibnamefont {Aidelsburger}},\ }\href@noop {} {\bibfield  {journal}
  {\bibinfo  {journal} {Physical Review X}\ }\textbf {\bibinfo {volume} {10}},\
  \bibinfo {pages} {011030} (\bibinfo {year} {2020})}\BibitemShut {NoStop}%
\bibitem [{\citenamefont {Pr{\"u}fer}\ \emph {et~al.}(2018)\citenamefont
  {Pr{\"u}fer}, \citenamefont {Kunkel}, \citenamefont {Strobel}, \citenamefont
  {Lannig}, \citenamefont {Linnemann}, \citenamefont {Schmied}, \citenamefont
  {Berges}, \citenamefont {Gasenzer},\ and\ \citenamefont
  {Oberthaler}}]{prufer2018observation}%
  \BibitemOpen
  \bibfield  {author} {\bibinfo {author} {\bibfnamefont {M.}~\bibnamefont
  {Pr{\"u}fer}}, \bibinfo {author} {\bibfnamefont {P.}~\bibnamefont {Kunkel}},
  \bibinfo {author} {\bibfnamefont {H.}~\bibnamefont {Strobel}}, \bibinfo
  {author} {\bibfnamefont {S.}~\bibnamefont {Lannig}}, \bibinfo {author}
  {\bibfnamefont {D.}~\bibnamefont {Linnemann}}, \bibinfo {author}
  {\bibfnamefont {C.-M.}\ \bibnamefont {Schmied}}, \bibinfo {author}
  {\bibfnamefont {J.}~\bibnamefont {Berges}}, \bibinfo {author} {\bibfnamefont
  {T.}~\bibnamefont {Gasenzer}}, \ and\ \bibinfo {author} {\bibfnamefont
  {M.~K.}\ \bibnamefont {Oberthaler}},\ }\href@noop {} {\bibfield  {journal}
  {\bibinfo  {journal} {Nature}\ }\textbf {\bibinfo {volume} {563}},\ \bibinfo
  {pages} {217} (\bibinfo {year} {2018})}\BibitemShut {NoStop}%
\bibitem [{\citenamefont {Erne}\ \emph {et~al.}(2018)\citenamefont {Erne},
  \citenamefont {B{\"u}cker}, \citenamefont {Gasenzer}, \citenamefont
  {Berges},\ and\ \citenamefont {Schmiedmayer}}]{erne2018universal}%
  \BibitemOpen
  \bibfield  {author} {\bibinfo {author} {\bibfnamefont {S.}~\bibnamefont
  {Erne}}, \bibinfo {author} {\bibfnamefont {R.}~\bibnamefont {B{\"u}cker}},
  \bibinfo {author} {\bibfnamefont {T.}~\bibnamefont {Gasenzer}}, \bibinfo
  {author} {\bibfnamefont {J.}~\bibnamefont {Berges}}, \ and\ \bibinfo {author}
  {\bibfnamefont {J.}~\bibnamefont {Schmiedmayer}},\ }\href@noop {} {\bibfield
  {journal} {\bibinfo  {journal} {Nature}\ }\textbf {\bibinfo {volume} {563}},\
  \bibinfo {pages} {225} (\bibinfo {year} {2018})}\BibitemShut {NoStop}%
\bibitem [{\citenamefont {Eigen}\ \emph {et~al.}(2018)\citenamefont {Eigen},
  \citenamefont {Glidden}, \citenamefont {Lopes}, \citenamefont {Cornell},
  \citenamefont {Smith},\ and\ \citenamefont
  {Hadzibabic}}]{eigen2018universal}%
  \BibitemOpen
  \bibfield  {author} {\bibinfo {author} {\bibfnamefont {C.}~\bibnamefont
  {Eigen}}, \bibinfo {author} {\bibfnamefont {J.~A.}\ \bibnamefont {Glidden}},
  \bibinfo {author} {\bibfnamefont {R.}~\bibnamefont {Lopes}}, \bibinfo
  {author} {\bibfnamefont {E.~A.}\ \bibnamefont {Cornell}}, \bibinfo {author}
  {\bibfnamefont {R.~P.}\ \bibnamefont {Smith}}, \ and\ \bibinfo {author}
  {\bibfnamefont {Z.}~\bibnamefont {Hadzibabic}},\ }\href@noop {} {\bibfield
  {journal} {\bibinfo  {journal} {Nature}\ }\textbf {\bibinfo {volume} {563}},\
  \bibinfo {pages} {221} (\bibinfo {year} {2018})}\BibitemShut {NoStop}%
\bibitem [{\citenamefont {Glidden}\ \emph {et~al.}(2021)\citenamefont
  {Glidden}, \citenamefont {Eigen}, \citenamefont {Dogra}, \citenamefont
  {Hilker}, \citenamefont {Smith},\ and\ \citenamefont
  {Hadzibabic}}]{glidden2021bidirectional}%
  \BibitemOpen
  \bibfield  {author} {\bibinfo {author} {\bibfnamefont {J.~A.}\ \bibnamefont
  {Glidden}}, \bibinfo {author} {\bibfnamefont {C.}~\bibnamefont {Eigen}},
  \bibinfo {author} {\bibfnamefont {L.~H.}\ \bibnamefont {Dogra}}, \bibinfo
  {author} {\bibfnamefont {T.~A.}\ \bibnamefont {Hilker}}, \bibinfo {author}
  {\bibfnamefont {R.~P.}\ \bibnamefont {Smith}}, \ and\ \bibinfo {author}
  {\bibfnamefont {Z.}~\bibnamefont {Hadzibabic}},\ }\href@noop {} {\bibfield
  {journal} {\bibinfo  {journal} {Nature Physics}\ }\textbf {\bibinfo {volume}
  {17}},\ \bibinfo {pages} {457} (\bibinfo {year} {2021})}\BibitemShut
  {NoStop}%
\bibitem [{\citenamefont {Navon}\ \emph {et~al.}(2015)\citenamefont {Navon},
  \citenamefont {Gaunt}, \citenamefont {Smith},\ and\ \citenamefont
  {Hadzibabic}}]{navon2015critical}%
  \BibitemOpen
  \bibfield  {author} {\bibinfo {author} {\bibfnamefont {N.}~\bibnamefont
  {Navon}}, \bibinfo {author} {\bibfnamefont {A.~L.}\ \bibnamefont {Gaunt}},
  \bibinfo {author} {\bibfnamefont {R.~P.}\ \bibnamefont {Smith}}, \ and\
  \bibinfo {author} {\bibfnamefont {Z.}~\bibnamefont {Hadzibabic}},\
  }\href@noop {} {\bibfield  {journal} {\bibinfo  {journal} {Science}\ }\textbf
  {\bibinfo {volume} {347}},\ \bibinfo {pages} {167} (\bibinfo {year}
  {2015})}\BibitemShut {NoStop}%
\bibitem [{\citenamefont {L{\"u}schen}\ \emph {et~al.}(2017)\citenamefont
  {L{\"u}schen}, \citenamefont {Bordia}, \citenamefont {Scherg}, \citenamefont
  {Alet}, \citenamefont {Altman}, \citenamefont {Schneider},\ and\
  \citenamefont {Bloch}}]{luschen2017observation}%
  \BibitemOpen
  \bibfield  {author} {\bibinfo {author} {\bibfnamefont {H.~P.}\ \bibnamefont
  {L{\"u}schen}}, \bibinfo {author} {\bibfnamefont {P.}~\bibnamefont {Bordia}},
  \bibinfo {author} {\bibfnamefont {S.}~\bibnamefont {Scherg}}, \bibinfo
  {author} {\bibfnamefont {F.}~\bibnamefont {Alet}}, \bibinfo {author}
  {\bibfnamefont {E.}~\bibnamefont {Altman}}, \bibinfo {author} {\bibfnamefont
  {U.}~\bibnamefont {Schneider}}, \ and\ \bibinfo {author} {\bibfnamefont
  {I.}~\bibnamefont {Bloch}},\ }\href@noop {} {\bibfield  {journal} {\bibinfo
  {journal} {Physical review letters}\ }\textbf {\bibinfo {volume} {119}},\
  \bibinfo {pages} {260401} (\bibinfo {year} {2017})}\BibitemShut {NoStop}%
\bibitem [{\citenamefont {Coleman}(1977)}]{coleman1977fate}%
  \BibitemOpen
  \bibfield  {author} {\bibinfo {author} {\bibfnamefont {S.}~\bibnamefont
  {Coleman}},\ }\href@noop {} {\bibfield  {journal} {\bibinfo  {journal}
  {Physical Review D}\ }\textbf {\bibinfo {volume} {15}},\ \bibinfo {pages}
  {2929} (\bibinfo {year} {1977})}\BibitemShut {NoStop}%
\bibitem [{\citenamefont {Mouchet}\ \emph {et~al.}(2001)\citenamefont
  {Mouchet}, \citenamefont {Miniatura}, \citenamefont {Kaiser}, \citenamefont
  {Gr{\'e}maud},\ and\ \citenamefont {Delande}}]{mouchet2001chaos}%
  \BibitemOpen
  \bibfield  {author} {\bibinfo {author} {\bibfnamefont {A.}~\bibnamefont
  {Mouchet}}, \bibinfo {author} {\bibfnamefont {C.}~\bibnamefont {Miniatura}},
  \bibinfo {author} {\bibfnamefont {R.}~\bibnamefont {Kaiser}}, \bibinfo
  {author} {\bibfnamefont {B.}~\bibnamefont {Gr{\'e}maud}}, \ and\ \bibinfo
  {author} {\bibfnamefont {D.}~\bibnamefont {Delande}},\ }\href@noop {}
  {\bibfield  {journal} {\bibinfo  {journal} {Physical Review E}\ }\textbf
  {\bibinfo {volume} {64}},\ \bibinfo {pages} {016221} (\bibinfo {year}
  {2001})}\BibitemShut {NoStop}%
\bibitem [{\citenamefont {Mehboudi}\ \emph {et~al.}(2019)\citenamefont
  {Mehboudi}, \citenamefont {Lampo}, \citenamefont {Charalambous},
  \citenamefont {Correa}, \citenamefont {Garc{\'\i}a-March},\ and\
  \citenamefont {Lewenstein}}]{mehboudi2019using}%
  \BibitemOpen
  \bibfield  {author} {\bibinfo {author} {\bibfnamefont {M.}~\bibnamefont
  {Mehboudi}}, \bibinfo {author} {\bibfnamefont {A.}~\bibnamefont {Lampo}},
  \bibinfo {author} {\bibfnamefont {C.}~\bibnamefont {Charalambous}}, \bibinfo
  {author} {\bibfnamefont {L.~A.}\ \bibnamefont {Correa}}, \bibinfo {author}
  {\bibfnamefont {M.~{\'A}.}\ \bibnamefont {Garc{\'\i}a-March}}, \ and\
  \bibinfo {author} {\bibfnamefont {M.}~\bibnamefont {Lewenstein}},\
  }\href@noop {} {\bibfield  {journal} {\bibinfo  {journal} {Physical review
  letters}\ }\textbf {\bibinfo {volume} {122}},\ \bibinfo {pages} {030403}
  (\bibinfo {year} {2019})}\BibitemShut {NoStop}%
\bibitem [{\citenamefont {Hu}\ \emph {et~al.}(2016)\citenamefont {Hu},
  \citenamefont {Van~de Graaff}, \citenamefont {Kedar}, \citenamefont {Corson},
  \citenamefont {Cornell},\ and\ \citenamefont {Jin}}]{hu2016bose}%
  \BibitemOpen
  \bibfield  {author} {\bibinfo {author} {\bibfnamefont {M.-G.}\ \bibnamefont
  {Hu}}, \bibinfo {author} {\bibfnamefont {M.~J.}\ \bibnamefont {Van~de
  Graaff}}, \bibinfo {author} {\bibfnamefont {D.}~\bibnamefont {Kedar}},
  \bibinfo {author} {\bibfnamefont {J.~P.}\ \bibnamefont {Corson}}, \bibinfo
  {author} {\bibfnamefont {E.~A.}\ \bibnamefont {Cornell}}, \ and\ \bibinfo
  {author} {\bibfnamefont {D.~S.}\ \bibnamefont {Jin}},\ }\href@noop {}
  {\bibfield  {journal} {\bibinfo  {journal} {Physical review letters}\
  }\textbf {\bibinfo {volume} {117}},\ \bibinfo {pages} {055301} (\bibinfo
  {year} {2016})}\BibitemShut {NoStop}%
\bibitem [{\citenamefont {J{\o}rgensen}\ \emph {et~al.}(2016)\citenamefont
  {J{\o}rgensen}, \citenamefont {Wacker}, \citenamefont {Skalmstang},
  \citenamefont {Parish}, \citenamefont {Levinsen}, \citenamefont
  {Christensen}, \citenamefont {Bruun},\ and\ \citenamefont
  {Arlt}}]{jorgensen2016observation}%
  \BibitemOpen
  \bibfield  {author} {\bibinfo {author} {\bibfnamefont {N.~B.}\ \bibnamefont
  {J{\o}rgensen}}, \bibinfo {author} {\bibfnamefont {L.}~\bibnamefont
  {Wacker}}, \bibinfo {author} {\bibfnamefont {K.~T.}\ \bibnamefont
  {Skalmstang}}, \bibinfo {author} {\bibfnamefont {M.~M.}\ \bibnamefont
  {Parish}}, \bibinfo {author} {\bibfnamefont {J.}~\bibnamefont {Levinsen}},
  \bibinfo {author} {\bibfnamefont {R.~S.}\ \bibnamefont {Christensen}},
  \bibinfo {author} {\bibfnamefont {G.~M.}\ \bibnamefont {Bruun}}, \ and\
  \bibinfo {author} {\bibfnamefont {J.~J.}\ \bibnamefont {Arlt}},\ }\href@noop
  {} {\bibfield  {journal} {\bibinfo  {journal} {Physical review letters}\
  }\textbf {\bibinfo {volume} {117}},\ \bibinfo {pages} {055302} (\bibinfo
  {year} {2016})}\BibitemShut {NoStop}%
\bibitem [{\citenamefont {Rentrop}\ \emph {et~al.}(2016)\citenamefont
  {Rentrop}, \citenamefont {Trautmann}, \citenamefont {Olivares}, \citenamefont
  {Jendrzejewski}, \citenamefont {Komnik},\ and\ \citenamefont
  {Oberthaler}}]{rentrop2016observation}%
  \BibitemOpen
  \bibfield  {author} {\bibinfo {author} {\bibfnamefont {T.}~\bibnamefont
  {Rentrop}}, \bibinfo {author} {\bibfnamefont {A.}~\bibnamefont {Trautmann}},
  \bibinfo {author} {\bibfnamefont {F.}~\bibnamefont {Olivares}}, \bibinfo
  {author} {\bibfnamefont {F.}~\bibnamefont {Jendrzejewski}}, \bibinfo {author}
  {\bibfnamefont {A.}~\bibnamefont {Komnik}}, \ and\ \bibinfo {author}
  {\bibfnamefont {M.~K.}\ \bibnamefont {Oberthaler}},\ }\href@noop {}
  {\bibfield  {journal} {\bibinfo  {journal} {Physical Review X}\ }\textbf
  {\bibinfo {volume} {6}},\ \bibinfo {pages} {041041} (\bibinfo {year}
  {2016})}\BibitemShut {NoStop}%
\bibitem [{\citenamefont {Aidelsburger}\ \emph {et~al.}(2022)\citenamefont
  {Aidelsburger}, \citenamefont {Barbiero}, \citenamefont {Bermudez},
  \citenamefont {Chanda}, \citenamefont {Dauphin}, \citenamefont
  {Gonz{\'a}lez-Cuadra}, \citenamefont {Grzybowski}, \citenamefont {Hands},
  \citenamefont {Jendrzejewski}, \citenamefont {J{\"u}nemann} \emph
  {et~al.}}]{aidelsburger2022cold}%
  \BibitemOpen
  \bibfield  {author} {\bibinfo {author} {\bibfnamefont {M.}~\bibnamefont
  {Aidelsburger}}, \bibinfo {author} {\bibfnamefont {L.}~\bibnamefont
  {Barbiero}}, \bibinfo {author} {\bibfnamefont {A.}~\bibnamefont {Bermudez}},
  \bibinfo {author} {\bibfnamefont {T.}~\bibnamefont {Chanda}}, \bibinfo
  {author} {\bibfnamefont {A.}~\bibnamefont {Dauphin}}, \bibinfo {author}
  {\bibfnamefont {D.}~\bibnamefont {Gonz{\'a}lez-Cuadra}}, \bibinfo {author}
  {\bibfnamefont {P.~R.}\ \bibnamefont {Grzybowski}}, \bibinfo {author}
  {\bibfnamefont {S.}~\bibnamefont {Hands}}, \bibinfo {author} {\bibfnamefont
  {F.}~\bibnamefont {Jendrzejewski}}, \bibinfo {author} {\bibfnamefont
  {J.}~\bibnamefont {J{\"u}nemann}},  \emph {et~al.},\ }\href@noop {}
  {\bibfield  {journal} {\bibinfo  {journal} {Philosophical Transactions of the
  Royal Society A}\ }\textbf {\bibinfo {volume} {380}},\ \bibinfo {pages}
  {20210064} (\bibinfo {year} {2022})}\BibitemShut {NoStop}%
\bibitem [{\citenamefont {Li}\ \emph {et~al.}(2015)\citenamefont {Li},
  \citenamefont {Zhu}, \citenamefont {He}, \citenamefont {Wang}, \citenamefont
  {Guo}, \citenamefont {Xu}, \citenamefont {Zhang},\ and\ \citenamefont
  {Wang}}]{Dajun2015}%
  \BibitemOpen
  \bibfield  {author} {\bibinfo {author} {\bibfnamefont {X.}~\bibnamefont
  {Li}}, \bibinfo {author} {\bibfnamefont {B.}~\bibnamefont {Zhu}}, \bibinfo
  {author} {\bibfnamefont {X.}~\bibnamefont {He}}, \bibinfo {author}
  {\bibfnamefont {F.}~\bibnamefont {Wang}}, \bibinfo {author} {\bibfnamefont
  {M.}~\bibnamefont {Guo}}, \bibinfo {author} {\bibfnamefont {Z.-F.}\
  \bibnamefont {Xu}}, \bibinfo {author} {\bibfnamefont {S.}~\bibnamefont
  {Zhang}}, \ and\ \bibinfo {author} {\bibfnamefont {D.}~\bibnamefont {Wang}},\
  }\href {\doibase 10.1103/PhysRevLett.114.255301} {\bibfield  {journal}
  {\bibinfo  {journal} {Phys. Rev. Lett.}\ }\textbf {\bibinfo {volume} {114}},\
  \bibinfo {pages} {255301} (\bibinfo {year} {2015})}\BibitemShut {NoStop}%
\bibitem [{\citenamefont {Mil}\ \emph {et~al.}(2020)\citenamefont {Mil},
  \citenamefont {Zache}, \citenamefont {Hegde}, \citenamefont {Xia},
  \citenamefont {Bhatt}, \citenamefont {Oberthaler}, \citenamefont {Hauke},
  \citenamefont {Berges},\ and\ \citenamefont
  {Jendrzejewski}}]{mil2020scalable}%
  \BibitemOpen
  \bibfield  {author} {\bibinfo {author} {\bibfnamefont {A.}~\bibnamefont
  {Mil}}, \bibinfo {author} {\bibfnamefont {T.~V.}\ \bibnamefont {Zache}},
  \bibinfo {author} {\bibfnamefont {A.}~\bibnamefont {Hegde}}, \bibinfo
  {author} {\bibfnamefont {A.}~\bibnamefont {Xia}}, \bibinfo {author}
  {\bibfnamefont {R.~P.}\ \bibnamefont {Bhatt}}, \bibinfo {author}
  {\bibfnamefont {M.~K.}\ \bibnamefont {Oberthaler}}, \bibinfo {author}
  {\bibfnamefont {P.}~\bibnamefont {Hauke}}, \bibinfo {author} {\bibfnamefont
  {J.}~\bibnamefont {Berges}}, \ and\ \bibinfo {author} {\bibfnamefont
  {F.}~\bibnamefont {Jendrzejewski}},\ }\href@noop {} {\bibfield  {journal}
  {\bibinfo  {journal} {Science}\ }\textbf {\bibinfo {volume} {367}},\ \bibinfo
  {pages} {1128} (\bibinfo {year} {2020})}\BibitemShut {NoStop}%
\bibitem [{\citenamefont {Dai}\ \emph {et~al.}(2017)\citenamefont {Dai},
  \citenamefont {Yang}, \citenamefont {Reingruber}, \citenamefont {Sun},
  \citenamefont {Xu}, \citenamefont {Chen}, \citenamefont {Yuan},\ and\
  \citenamefont {Pan}}]{dai2017four}%
  \BibitemOpen
  \bibfield  {author} {\bibinfo {author} {\bibfnamefont {H.-N.}\ \bibnamefont
  {Dai}}, \bibinfo {author} {\bibfnamefont {B.}~\bibnamefont {Yang}}, \bibinfo
  {author} {\bibfnamefont {A.}~\bibnamefont {Reingruber}}, \bibinfo {author}
  {\bibfnamefont {H.}~\bibnamefont {Sun}}, \bibinfo {author} {\bibfnamefont
  {X.-F.}\ \bibnamefont {Xu}}, \bibinfo {author} {\bibfnamefont {Y.-A.}\
  \bibnamefont {Chen}}, \bibinfo {author} {\bibfnamefont {Z.-S.}\ \bibnamefont
  {Yuan}}, \ and\ \bibinfo {author} {\bibfnamefont {J.-W.}\ \bibnamefont
  {Pan}},\ }\href@noop {} {\bibfield  {journal} {\bibinfo  {journal} {Nature
  Physics}\ }\textbf {\bibinfo {volume} {13}},\ \bibinfo {pages} {1195}
  (\bibinfo {year} {2017})}\BibitemShut {NoStop}%
\bibitem [{\citenamefont {Schweizer}\ \emph {et~al.}(2019)\citenamefont
  {Schweizer}, \citenamefont {Grusdt}, \citenamefont {Berngruber},
  \citenamefont {Barbiero}, \citenamefont {Demler}, \citenamefont {Goldman},
  \citenamefont {Bloch},\ and\ \citenamefont
  {Aidelsburger}}]{schweizer2019floquet}%
  \BibitemOpen
  \bibfield  {author} {\bibinfo {author} {\bibfnamefont {C.}~\bibnamefont
  {Schweizer}}, \bibinfo {author} {\bibfnamefont {F.}~\bibnamefont {Grusdt}},
  \bibinfo {author} {\bibfnamefont {M.}~\bibnamefont {Berngruber}}, \bibinfo
  {author} {\bibfnamefont {L.}~\bibnamefont {Barbiero}}, \bibinfo {author}
  {\bibfnamefont {E.}~\bibnamefont {Demler}}, \bibinfo {author} {\bibfnamefont
  {N.}~\bibnamefont {Goldman}}, \bibinfo {author} {\bibfnamefont
  {I.}~\bibnamefont {Bloch}}, \ and\ \bibinfo {author} {\bibfnamefont
  {M.}~\bibnamefont {Aidelsburger}},\ }\href@noop {} {\bibfield  {journal}
  {\bibinfo  {journal} {Nature Physics}\ }\textbf {\bibinfo {volume} {15}},\
  \bibinfo {pages} {1168} (\bibinfo {year} {2019})}\BibitemShut {NoStop}%
\bibitem [{\citenamefont {Stamper-Kurn}\ and\ \citenamefont
  {Ueda}(2013)}]{stamper2013spinor}%
  \BibitemOpen
  \bibfield  {author} {\bibinfo {author} {\bibfnamefont {D.~M.}\ \bibnamefont
  {Stamper-Kurn}}\ and\ \bibinfo {author} {\bibfnamefont {M.}~\bibnamefont
  {Ueda}},\ }\href@noop {} {\bibfield  {journal} {\bibinfo  {journal} {Reviews
  of Modern Physics}\ }\textbf {\bibinfo {volume} {85}},\ \bibinfo {pages}
  {1191} (\bibinfo {year} {2013})}\BibitemShut {NoStop}%
\bibitem [{Note1()}]{Note1}%
  \BibitemOpen
  \bibinfo {note} {Definitions of imbalances for Na are analogous.}\BibitemShut
  {Stop}%
\bibitem [{\citenamefont {Garc{\'\i}a-Palacios}\ and\ \citenamefont
  {L{\'a}zaro}(1998)}]{garcia1998langevin}%
  \BibitemOpen
  \bibfield  {author} {\bibinfo {author} {\bibfnamefont {J.~L.}\ \bibnamefont
  {Garc{\'\i}a-Palacios}}\ and\ \bibinfo {author} {\bibfnamefont {F.~J.}\
  \bibnamefont {L{\'a}zaro}},\ }\href@noop {} {\bibfield  {journal} {\bibinfo
  {journal} {Physical Review B}\ }\textbf {\bibinfo {volume} {58}},\ \bibinfo
  {pages} {14937} (\bibinfo {year} {1998})}\BibitemShut {NoStop}%
\bibitem [{\citenamefont {Zache}(2020)}]{torsten2020}%
  \BibitemOpen
  \bibfield  {author} {\bibinfo {author} {\bibfnamefont {T.~V.}\ \bibnamefont
  {Zache}},\ }\href {\doibase 10.11588/heidok.00028536} {\bibfield  {journal}
  {\bibinfo  {journal} {Doctoral thesis}\ } (\bibinfo {year} {2020}),\
  10.11588/heidok.00028536}\BibitemShut {NoStop}%
\bibitem [{\citenamefont {Schweigler}\ \emph {et~al.}(2017)\citenamefont
  {Schweigler}, \citenamefont {Kasper}, \citenamefont {Erne}, \citenamefont
  {Mazets}, \citenamefont {Rauer}, \citenamefont {Cataldini}, \citenamefont
  {Langen}, \citenamefont {Gasenzer}, \citenamefont {Berges},\ and\
  \citenamefont {Schmiedmayer}}]{schweigler2017experimental}%
  \BibitemOpen
  \bibfield  {author} {\bibinfo {author} {\bibfnamefont {T.}~\bibnamefont
  {Schweigler}}, \bibinfo {author} {\bibfnamefont {V.}~\bibnamefont {Kasper}},
  \bibinfo {author} {\bibfnamefont {S.}~\bibnamefont {Erne}}, \bibinfo {author}
  {\bibfnamefont {I.}~\bibnamefont {Mazets}}, \bibinfo {author} {\bibfnamefont
  {B.}~\bibnamefont {Rauer}}, \bibinfo {author} {\bibfnamefont
  {F.}~\bibnamefont {Cataldini}}, \bibinfo {author} {\bibfnamefont
  {T.}~\bibnamefont {Langen}}, \bibinfo {author} {\bibfnamefont
  {T.}~\bibnamefont {Gasenzer}}, \bibinfo {author} {\bibfnamefont
  {J.}~\bibnamefont {Berges}}, \ and\ \bibinfo {author} {\bibfnamefont
  {J.}~\bibnamefont {Schmiedmayer}},\ }\href@noop {} {\bibfield  {journal}
  {\bibinfo  {journal} {Nature}\ }\textbf {\bibinfo {volume} {545}},\ \bibinfo
  {pages} {323} (\bibinfo {year} {2017})}\BibitemShut {NoStop}%
\bibitem [{\citenamefont {Zache}\ \emph {et~al.}(2020)\citenamefont {Zache},
  \citenamefont {Schweigler}, \citenamefont {Erne}, \citenamefont
  {Schmiedmayer},\ and\ \citenamefont {Berges}}]{zache2020extracting}%
  \BibitemOpen
  \bibfield  {author} {\bibinfo {author} {\bibfnamefont {T.~V.}\ \bibnamefont
  {Zache}}, \bibinfo {author} {\bibfnamefont {T.}~\bibnamefont {Schweigler}},
  \bibinfo {author} {\bibfnamefont {S.}~\bibnamefont {Erne}}, \bibinfo {author}
  {\bibfnamefont {J.}~\bibnamefont {Schmiedmayer}}, \ and\ \bibinfo {author}
  {\bibfnamefont {J.}~\bibnamefont {Berges}},\ }\href@noop {} {\bibfield
  {journal} {\bibinfo  {journal} {Physical Review X}\ }\textbf {\bibinfo
  {volume} {10}},\ \bibinfo {pages} {011020} (\bibinfo {year}
  {2020})}\BibitemShut {NoStop}%
\bibitem [{\citenamefont {Rispoli}\ \emph {et~al.}(2019)\citenamefont
  {Rispoli}, \citenamefont {Lukin}, \citenamefont {Schittko}, \citenamefont
  {Kim}, \citenamefont {Tai}, \citenamefont {L{\'e}onard},\ and\ \citenamefont
  {Greiner}}]{rispoli2019quantum}%
  \BibitemOpen
  \bibfield  {author} {\bibinfo {author} {\bibfnamefont {M.}~\bibnamefont
  {Rispoli}}, \bibinfo {author} {\bibfnamefont {A.}~\bibnamefont {Lukin}},
  \bibinfo {author} {\bibfnamefont {R.}~\bibnamefont {Schittko}}, \bibinfo
  {author} {\bibfnamefont {S.}~\bibnamefont {Kim}}, \bibinfo {author}
  {\bibfnamefont {M.~E.}\ \bibnamefont {Tai}}, \bibinfo {author} {\bibfnamefont
  {J.}~\bibnamefont {L{\'e}onard}}, \ and\ \bibinfo {author} {\bibfnamefont
  {M.}~\bibnamefont {Greiner}},\ }\href@noop {} {\bibfield  {journal} {\bibinfo
   {journal} {Nature}\ }\textbf {\bibinfo {volume} {573}},\ \bibinfo {pages}
  {385} (\bibinfo {year} {2019})}\BibitemShut {NoStop}%
\bibitem [{\citenamefont {Pr{\"u}fer}\ \emph {et~al.}(2020)\citenamefont
  {Pr{\"u}fer}, \citenamefont {Zache}, \citenamefont {Kunkel}, \citenamefont
  {Lannig}, \citenamefont {Bonnin}, \citenamefont {Strobel}, \citenamefont
  {Berges},\ and\ \citenamefont {Oberthaler}}]{prufer2020experimental}%
  \BibitemOpen
  \bibfield  {author} {\bibinfo {author} {\bibfnamefont {M.}~\bibnamefont
  {Pr{\"u}fer}}, \bibinfo {author} {\bibfnamefont {T.~V.}\ \bibnamefont
  {Zache}}, \bibinfo {author} {\bibfnamefont {P.}~\bibnamefont {Kunkel}},
  \bibinfo {author} {\bibfnamefont {S.}~\bibnamefont {Lannig}}, \bibinfo
  {author} {\bibfnamefont {A.}~\bibnamefont {Bonnin}}, \bibinfo {author}
  {\bibfnamefont {H.}~\bibnamefont {Strobel}}, \bibinfo {author} {\bibfnamefont
  {J.}~\bibnamefont {Berges}}, \ and\ \bibinfo {author} {\bibfnamefont {M.~K.}\
  \bibnamefont {Oberthaler}},\ }\href@noop {} {\bibfield  {journal} {\bibinfo
  {journal} {Nature Physics}\ }\textbf {\bibinfo {volume} {16}},\ \bibinfo
  {pages} {1012} (\bibinfo {year} {2020})}\BibitemShut {NoStop}%
\bibitem [{\citenamefont {Ott}\ \emph {et~al.}(2021)\citenamefont {Ott},
  \citenamefont {Zache}, \citenamefont {Jendrzejewski},\ and\ \citenamefont
  {Berges}}]{ott2021scalable}%
  \BibitemOpen
  \bibfield  {author} {\bibinfo {author} {\bibfnamefont {R.}~\bibnamefont
  {Ott}}, \bibinfo {author} {\bibfnamefont {T.~V.}\ \bibnamefont {Zache}},
  \bibinfo {author} {\bibfnamefont {F.}~\bibnamefont {Jendrzejewski}}, \ and\
  \bibinfo {author} {\bibfnamefont {J.}~\bibnamefont {Berges}},\ }\href@noop {}
  {\bibfield  {journal} {\bibinfo  {journal} {Physical Review Letters}\
  }\textbf {\bibinfo {volume} {127}},\ \bibinfo {pages} {130504} (\bibinfo
  {year} {2021})}\BibitemShut {NoStop}%
\bibitem [{\citenamefont {Surace}\ \emph {et~al.}(2020)\citenamefont {Surace},
  \citenamefont {Mazza}, \citenamefont {Giudici}, \citenamefont {Lerose},
  \citenamefont {Gambassi},\ and\ \citenamefont
  {Dalmonte}}]{surace2020lattice}%
  \BibitemOpen
  \bibfield  {author} {\bibinfo {author} {\bibfnamefont {F.~M.}\ \bibnamefont
  {Surace}}, \bibinfo {author} {\bibfnamefont {P.~P.}\ \bibnamefont {Mazza}},
  \bibinfo {author} {\bibfnamefont {G.}~\bibnamefont {Giudici}}, \bibinfo
  {author} {\bibfnamefont {A.}~\bibnamefont {Lerose}}, \bibinfo {author}
  {\bibfnamefont {A.}~\bibnamefont {Gambassi}}, \ and\ \bibinfo {author}
  {\bibfnamefont {M.}~\bibnamefont {Dalmonte}},\ }\href@noop {} {\bibfield
  {journal} {\bibinfo  {journal} {Physical Review X}\ }\textbf {\bibinfo
  {volume} {10}},\ \bibinfo {pages} {021041} (\bibinfo {year}
  {2020})}\BibitemShut {NoStop}%
\bibitem [{\citenamefont {G{\"o}rg}\ \emph {et~al.}(2019)\citenamefont
  {G{\"o}rg}, \citenamefont {Sandholzer}, \citenamefont {Minguzzi},
  \citenamefont {Desbuquois}, \citenamefont {Messer},\ and\ \citenamefont
  {Esslinger}}]{gorg2019realization}%
  \BibitemOpen
  \bibfield  {author} {\bibinfo {author} {\bibfnamefont {F.}~\bibnamefont
  {G{\"o}rg}}, \bibinfo {author} {\bibfnamefont {K.}~\bibnamefont
  {Sandholzer}}, \bibinfo {author} {\bibfnamefont {J.}~\bibnamefont
  {Minguzzi}}, \bibinfo {author} {\bibfnamefont {R.}~\bibnamefont
  {Desbuquois}}, \bibinfo {author} {\bibfnamefont {M.}~\bibnamefont {Messer}},
  \ and\ \bibinfo {author} {\bibfnamefont {T.}~\bibnamefont {Esslinger}},\
  }\href@noop {} {\bibfield  {journal} {\bibinfo  {journal} {Nature Physics}\
  }\textbf {\bibinfo {volume} {15}},\ \bibinfo {pages} {1161} (\bibinfo {year}
  {2019})}\BibitemShut {NoStop}%
\bibitem [{\citenamefont {Martinez}\ \emph {et~al.}(2016)\citenamefont
  {Martinez}, \citenamefont {Muschik}, \citenamefont {Schindler}, \citenamefont
  {Nigg}, \citenamefont {Erhard}, \citenamefont {Heyl}, \citenamefont {Hauke},
  \citenamefont {Dalmonte}, \citenamefont {Monz}, \citenamefont {Zoller} \emph
  {et~al.}}]{martinez2016real}%
  \BibitemOpen
  \bibfield  {author} {\bibinfo {author} {\bibfnamefont {E.~A.}\ \bibnamefont
  {Martinez}}, \bibinfo {author} {\bibfnamefont {C.~A.}\ \bibnamefont
  {Muschik}}, \bibinfo {author} {\bibfnamefont {P.}~\bibnamefont {Schindler}},
  \bibinfo {author} {\bibfnamefont {D.}~\bibnamefont {Nigg}}, \bibinfo {author}
  {\bibfnamefont {A.}~\bibnamefont {Erhard}}, \bibinfo {author} {\bibfnamefont
  {M.}~\bibnamefont {Heyl}}, \bibinfo {author} {\bibfnamefont {P.}~\bibnamefont
  {Hauke}}, \bibinfo {author} {\bibfnamefont {M.}~\bibnamefont {Dalmonte}},
  \bibinfo {author} {\bibfnamefont {T.}~\bibnamefont {Monz}}, \bibinfo {author}
  {\bibfnamefont {P.}~\bibnamefont {Zoller}},  \emph {et~al.},\ }\href@noop {}
  {\bibfield  {journal} {\bibinfo  {journal} {Nature}\ }\textbf {\bibinfo
  {volume} {534}},\ \bibinfo {pages} {516} (\bibinfo {year}
  {2016})}\BibitemShut {NoStop}%
\bibitem [{\citenamefont {Klco}\ \emph {et~al.}(2018)\citenamefont {Klco},
  \citenamefont {Dumitrescu}, \citenamefont {McCaskey}, \citenamefont {Morris},
  \citenamefont {Pooser}, \citenamefont {Sanz}, \citenamefont {Solano},
  \citenamefont {Lougovski},\ and\ \citenamefont {Savage}}]{klco2018quantum}%
  \BibitemOpen
  \bibfield  {author} {\bibinfo {author} {\bibfnamefont {N.}~\bibnamefont
  {Klco}}, \bibinfo {author} {\bibfnamefont {E.~F.}\ \bibnamefont
  {Dumitrescu}}, \bibinfo {author} {\bibfnamefont {A.~J.}\ \bibnamefont
  {McCaskey}}, \bibinfo {author} {\bibfnamefont {T.~D.}\ \bibnamefont
  {Morris}}, \bibinfo {author} {\bibfnamefont {R.~C.}\ \bibnamefont {Pooser}},
  \bibinfo {author} {\bibfnamefont {M.}~\bibnamefont {Sanz}}, \bibinfo {author}
  {\bibfnamefont {E.}~\bibnamefont {Solano}}, \bibinfo {author} {\bibfnamefont
  {P.}~\bibnamefont {Lougovski}}, \ and\ \bibinfo {author} {\bibfnamefont
  {M.~J.}\ \bibnamefont {Savage}},\ }\href@noop {} {\bibfield  {journal}
  {\bibinfo  {journal} {Physical Review A}\ }\textbf {\bibinfo {volume} {98}},\
  \bibinfo {pages} {032331} (\bibinfo {year} {2018})}\BibitemShut {NoStop}%
\bibitem [{\citenamefont {Zhou}\ \emph {et~al.}(2021)\citenamefont {Zhou},
  \citenamefont {Su}, \citenamefont {Halimeh}, \citenamefont {Ott},
  \citenamefont {Sun}, \citenamefont {Hauke}, \citenamefont {Yang},
  \citenamefont {Yuan}, \citenamefont {Berges},\ and\ \citenamefont
  {Pan}}]{zhou2021thermalization}%
  \BibitemOpen
  \bibfield  {author} {\bibinfo {author} {\bibfnamefont {Z.-Y.}\ \bibnamefont
  {Zhou}}, \bibinfo {author} {\bibfnamefont {G.-X.}\ \bibnamefont {Su}},
  \bibinfo {author} {\bibfnamefont {J.~C.}\ \bibnamefont {Halimeh}}, \bibinfo
  {author} {\bibfnamefont {R.}~\bibnamefont {Ott}}, \bibinfo {author}
  {\bibfnamefont {H.}~\bibnamefont {Sun}}, \bibinfo {author} {\bibfnamefont
  {P.}~\bibnamefont {Hauke}}, \bibinfo {author} {\bibfnamefont
  {B.}~\bibnamefont {Yang}}, \bibinfo {author} {\bibfnamefont {Z.-S.}\
  \bibnamefont {Yuan}}, \bibinfo {author} {\bibfnamefont {J.}~\bibnamefont
  {Berges}}, \ and\ \bibinfo {author} {\bibfnamefont {J.-W.}\ \bibnamefont
  {Pan}},\ }\href@noop {} {\bibfield  {journal} {\bibinfo  {journal} {arXiv
  preprint arXiv:2107.13563}\ } (\bibinfo {year} {2021})}\BibitemShut {NoStop}%
\bibitem [{\citenamefont {Mildenberger}\ \emph {et~al.}(2022)\citenamefont
  {Mildenberger}, \citenamefont {Mruczkiewicz}, \citenamefont {Halimeh},
  \citenamefont {Jiang},\ and\ \citenamefont
  {Hauke}}]{mildenberger2022probing}%
  \BibitemOpen
  \bibfield  {author} {\bibinfo {author} {\bibfnamefont {J.}~\bibnamefont
  {Mildenberger}}, \bibinfo {author} {\bibfnamefont {W.}~\bibnamefont
  {Mruczkiewicz}}, \bibinfo {author} {\bibfnamefont {J.~C.}\ \bibnamefont
  {Halimeh}}, \bibinfo {author} {\bibfnamefont {Z.}~\bibnamefont {Jiang}}, \
  and\ \bibinfo {author} {\bibfnamefont {P.}~\bibnamefont {Hauke}},\
  }\href@noop {} {\bibfield  {journal} {\bibinfo  {journal} {arXiv preprint
  arXiv:2203.08905}\ } (\bibinfo {year} {2022})}\BibitemShut {NoStop}%
\end{thebibliography}%
\clearpage

\section{Supplemental Material}

\subsection{Model Hamiltonian}
\label{SM:Hamiltonian}
In this section, we motivate the choice of Hamiltonian~\eqref{eq:Hamiltonian} through the microscopic theory of $F=1$ spinor Bose gases \cite{stamper2013spinor}, see also Ref.~\cite{mil2020scalable}. We consider the microscopic Hamiltonian
\begin{align}
	H_{0} = \sum_{m_F}\int \mathrm{d}^{3}x \Bigg[ \hat{\psi}^\dagger_{x,m_F} \Big(-\frac{\nabla^2}{2m}&-\mu + V_{m_F}(x) \Big) \hat{\psi}_{x,m_F} \nonumber\\
	&+ : c_0 \hat{n}_x^2 + c_1 \hat{\mathbf{F}}_x^2: \Bigg],
\end{align}  
with atomic mass $m$, chemical potential $\mu$, and scattering constants $c_0$, $c_1$ for local interactions of density $\hat{n}_x= \sum_{m_F}\hat{\psi}^\dagger_{x,m_F}\hat{\psi}_{x,m_F}$ and spin $\hat{\mathbf{F}}_{x}= \sum_{m_F,m_F'}\hat{\psi}^\dagger_{x,m_F} \mathbf{f}_{m_Fm_F'}\hat{\psi}_{x,m_F'}$. Here, $\hat{\psi}_{x,m_F} = \hat{\psi}_{m_F}(x)$ are the bosonic field operators for hyperfine states $m_F=-1,0,1$ with commutation relations $[\hat{\psi}_{x,m_F},\hat{\psi}_{y,m_F'}^\dagger] = \delta_{m_Fm_F'}\delta(x-y)$, where $\mathbf{f}= (f^{x},f^{y},f^{z})$ are the spin-1 matrices, and the colon represents normal ordering of operators.

The atomic mixture combines two such Bose gases: Na and Li. They are coupled through an interaction Hamiltonian
\begin{align}
	\hat{H}_\mathrm{int} = c_0^{\mathrm{NaLi}} \hat{n}_x^\mathrm{Na} \hat{n}_x^\mathrm{Li} + c_1^{\mathrm{NaLi}} \hat{\mathbf{F}}_x^\mathrm{Na} \cdot \hat{\mathbf{F}}_x^\mathrm{Li}\; ,
\end{align}
such that the total Hamiltonian reads
\begin{align}
	\hat{H}_\mathrm{NaLi} = \hat{H}_{\mathrm{Na},0} + \hat{H}_{\mathrm{Li},0} + \hat{H}_\mathrm{int} \; .
\end{align} 
In the experiment the atoms are strongly confined in a crossed optical dipole trap, i.e. they are tightly bound by optical potentials to suppress spatial dynamics in the gas. The system may effectively be described by a model with a single spatial mode for each magnetic substate. Detuning the $m_F=-1$ subspace from the dynamics, the free quadratic part of the resulting effective Hamiltonian reads
\begin{align}
	\hat{H}_{\mathrm{NaLi},0} = \sum_{a,m_F=0,1} \epsilon_{a,m_F} \hat{b}^\dagger_{a,m_F}\hat{b}_{a,m_F} ,
\end{align} 
where $m_F$ sums are restricted to $0$ and $1$ in the following. The single-particle energies $\epsilon_{a,m_F}$ are set by the trapping frequencies, external fields, fluctuations of spatial modes and we absorb further terms which arise from reordering the interaction operators. For fixed total atom numbers and magnetization, it may be expressed in a rotating frame as
\begin{align}
    \hat{H}_{\mathrm{NaLi},0} &=\frac{\tilde{\Delta}}{2}\sum_{m_F}  (-1)^{m_F} \hat{b}^\dagger_{\mathrm{Li},m_F}\hat{b}_{\mathrm{Li},m_F}\nonumber\\
    &= \tilde{\Delta} \hat{L}_{\mathrm{Li},z} .
\end{align}
In this approximation, the interaction Hamiltonian is given by
\begin{align}
    \hat{H}_\mathrm{int} &= \sum_{a,m_F}\chi_{a,m_F} \hat{N}_{a,m_F}^2  \nonumber\\
    &+ \ \chi_{\mathrm{NaLi}}(\hat{N}_{\mathrm{Na},0}-\hat{N}_{\mathrm{Na},1})(\hat{N}_{\mathrm{Li},0}-\hat{N}_{\mathrm{Li},1})\nonumber\\
    &+ \lambda\left( \hat{b}^\dagger_{\mathrm{Na},0}\hat{b}^\dagger_{\mathrm{Li},1}\hat{b}_{\mathrm{Na},1}\hat{b}_{\mathrm{Li},0} + h.c.\right) \; ,
\end{align}
and all other combinations of operators have been absorbed into single-particle couplings. The effective interaction constants involve experimental details such as the scattering lengths, the atomic masses, magnetic fields and the spatial overlap of the condensates. We simplify the expression by using
\begin{align}
    \chi_{a,m_F}&\hat{N}_{a,m_F}^2 = \chi_{a,m_F} \left(\frac{\hat{M}_{a}}{2} + (-1)^{m_F}  \hat{L}_{a,z} \right)^2 \nonumber\\
    &= \chi_{a,m_F} \left( \frac{\hat{N}^2_{a}}{4} + \hat{L}_{a,z}^2 +  (-1)^{m_F}  \hat{N}_{a}\hat{L}_{a,z} \right) .
\end{align}
By collecting all terms, assuming conservation of total atom numbers and magnetization during the dynamics, and dropping irrelevant constants, we may write the Hamiltonian as in Eq.~\eqref{eq:Hamiltonian} with the identifications
\begin{align}
\label{SM:params_theory}
    \Delta &= \tilde{\Delta} + 4\chi_\mathrm{NaLi}\hat{M}+ 2\hat{M} (\chi_{\mathrm{Li},0}+\chi_{\mathrm{Li},1})\nonumber\\
    &+ (\chi_{\mathrm{Li},0}-\chi_{\mathrm{Li},1})\hat{N}_{\mathrm{Li}}\nonumber -(\chi_{\mathrm{Na},0}-\chi_{\mathrm{Na},1})\hat{N}_{\mathrm{Na}} , \\
    \chi &= \chi_{\mathrm{Na},0}+\chi_{\mathrm{Na},1}+\chi_{\mathrm{Li},0}+\chi_{\mathrm{Li},1} -4\chi_\mathrm{NaLi}.
\end{align}
For the chosen initial states, the magnetization is set by $\hat{L}_{\mathrm{Na},z}(t_0)- \hat{N}_\mathrm{Li}/2$. Neglecting fluctuations of the Li atom number in the model parameters, $N_\mathrm{Li}= \langle \hat{N}_\mathrm{Li}\rangle$ is absorbed into the parameters, and we get
\begin{align}
\label{app:Delta}
    \Delta = \Delta_0 + 2\Delta_L \hat{L}_{\mathrm{Na},z}(t_0) + \Delta_N \hat{N}_\mathrm{Na} ,
\end{align}
where, specifically, we defined
\begin{subequations}
\begin{align}
    \Delta_L&=  2\chi_\mathrm{NaLi}+  (\chi_{\mathrm{Li},0}+\chi_{\mathrm{Li},1})\,,\\
    \Delta_0&= \tilde{\Delta} -\Delta_L N_\mathrm{Li} \, ,\\
    \Delta_N&=  \chi_{\mathrm{Na},1}-\chi_{\mathrm{Na},0} \,.
\end{align}
\end{subequations}
In our numerical calculations we consider $\Delta$ including fluctuations of $\hat{L}_{\mathrm{Na},z}(t_0)$, and $\hat{N}_\mathrm{Na}$ in the initial state as given in \eqref{app:Delta}

\subsection{Model parameters}
In this section, we determine the coupling parameters entering our model in the Hamiltonian given by Eq.~\eqref{eq:Hamiltonian}. To achieve this, we post-select the experimental data to analyze the dependence of model parameters on the number of Na atoms in the atomic cloud. We assume that the atom number of Na has a relatively bigger impact on the parameters compared to Li, as it is significantly higher, and hence we only consider atom number fluctuations of Na in this section. We focus our analysis on the data-set used in \Fig{fig:Imbalance}.

For the considered data set we measured the atom number of Na at $t=30$ms as $N_\mathrm{Na} = (376.5 \pm 20.4)\times 10^3 $, where the atom number fluctuation $\sigma = 20.4\times 10^3$ is the standard deviation of the distribution of all experimental realizations. We split the data set into six ``bands" of different mean atom numbers but fixed width set by $\sigma/2$, and selectively extract the lithium imbalance values $\eta_\mathrm{Li}$. We then fit numerical calculations to the data to extract parameters for the different total atom numbers. 

\begin{figure}[h!]
    \includegraphics[width=0.48\textwidth]{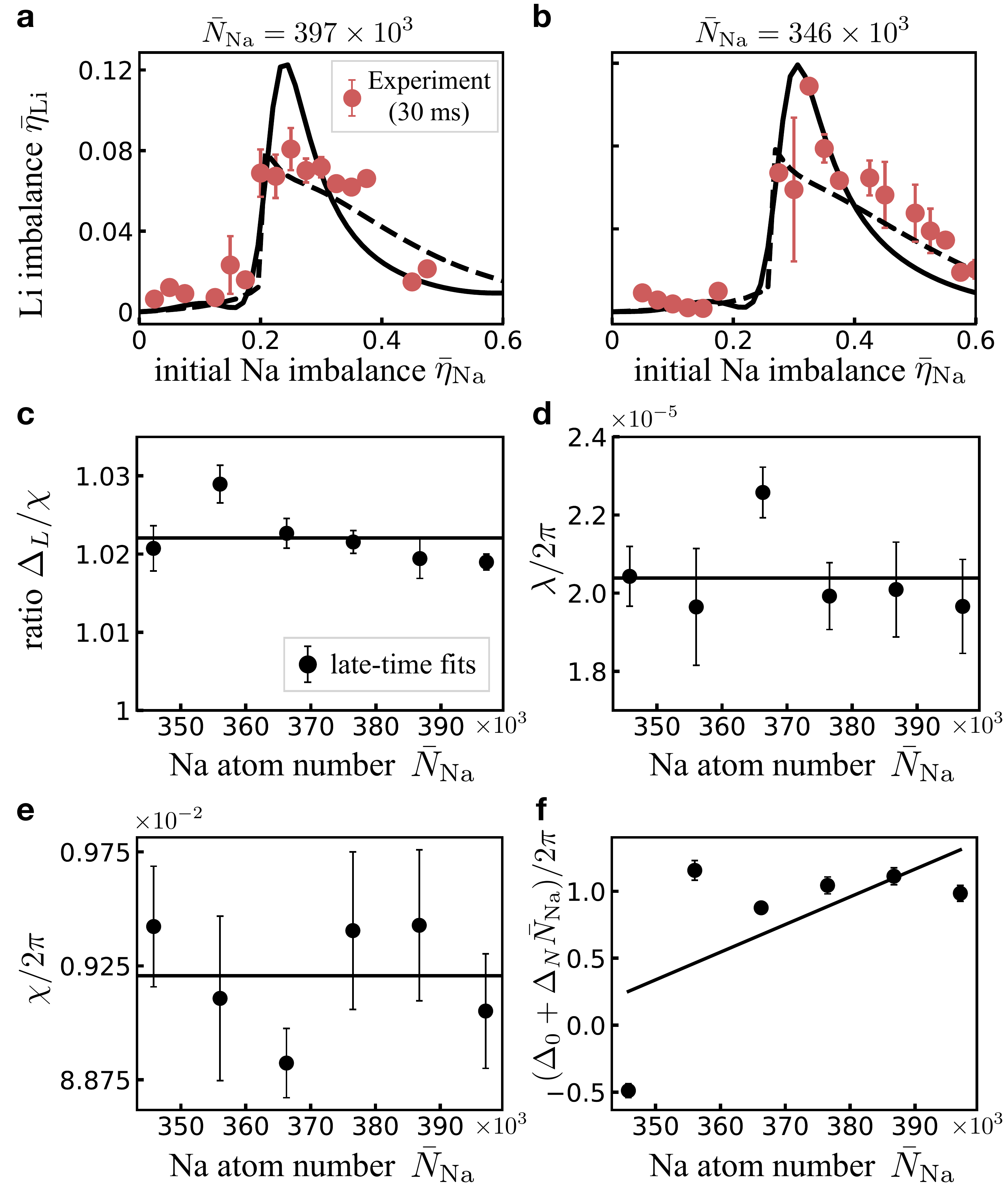}
    \caption{\textbf{Model params:} (\textbf{a}, \textbf{b}) We fit the data with the simulation result for $t=150$ms (dashed) for two different mean Na atom numbers $\bar{N}_\mathrm{Na}$ (as a comparison we also show the numerical result for $t=30$ms (solid)). (\textbf{c}-\textbf{f}) The parameters we obtain from fitting six different mean atom numbers are fitted to the model parameters.}
    \label{fig:post_sel}
\end{figure}

We extract the atom number dependence of parameters by fitting the data to a mean-field calculation of our model. As shown in \Fig{fig:Dynamics}, the observed Li imbalance at evolution time $t=30$ms is an estimate for stationary states at late times. Using this reasoning, we extract an estimate for our model parameters by fitting the data to the numerical calculation at $t=150$ms with a damping rate $\gamma = 2.2\times 10^{-3}$Hz. For example, two such fits corresponding to the mean atom numbers $\bar{N}_\mathrm{Na}= 397\times 10^3$ and $\bar{N}_\mathrm{Na}= 346 \times 10^3$ are shown in \Fig{fig:post_sel}a. Matching the resonance requires a fine-tuning of $\chi$ and $\Delta$. In \Fig{fig:post_sel} we show parameter sets obtained from these fits. Panel b indicates that $\chi$ and $\Delta_L$ are approximately equal on the percentage level. Further, both the ratio $\Delta_L/\chi$ as well as the rate of the spin-changing collisions only weakly depend on the Na atom number $\bar{N}_\mathrm{Na}$ in the considered range, while $\chi$ and $\Delta_L$ show a stronger dependence. From the procedure explained above, we obtain $\lambda/2\pi= 2.038\times 10^{-5}$Hz, $\chi/2\pi=0.0092$Hz, $\Delta_L/2\pi = 0.00941$Hz, $\Delta_N/2\pi = -1.298\times 10^{-4}$Hz and $\Delta_0/2\pi=43.30$Hz, by fitting the values in \Fig{fig:post_sel} with our model parameters. The similar values $\Delta_L\approx \chi$ as well as the relatively smaller $|\Delta_N|\ll \chi$ suggest that the dynamics of density interactions are dominated by the Li scattering constants $\chi_{\mathrm{L,m_F}}$.  

We use these parameters as a starting point to match both the mean and fluctuation data for both \Fig{fig:Dynamics} and \Fig{fig:Imbalance}, and obtain
    $\lambda/2\pi = 1.8\times 10^{5}$Hz, $\chi/2\pi =0.00943$Hz, $\Delta_L/2\pi = 0.00961$Hz, $\Delta_0/2\pi=-73.81$Hz, $\Delta_N/2\pi = 2.108\times 10^{-4}$Hz, and $\gamma = 1.8\times 10^{-3}$Hz.

\subsection{Numerical evolution}
In our numerical real-time calculations we compute the classical evolution equations for an ensemble of initial states. The ensemble is characterized by Gaussian distributions of total atom numbers and magnetization. For the distribution of atom numbers we use the respectively measured mean values and standard deviations as $N_\mathrm{Na} = (313.2\pm 18.3)\times 10^3$, $N_\mathrm{Li} = (34.6\pm 3.2)\times 10^3$ for \Fig{fig:Dynamics}, and $N_\mathrm{Na} = (376.5\pm 20.4)\times 10^3$, $N_\mathrm{Li} = (29.0\pm 2.3)\times 10^3$ for \Fig{fig:Imbalance}. Further, we first consider the atoms in their respective $\ket{F=1,m_F=1}$ state as also prepared in the experiments. Here, we take into account the (Gaussian) quantum fluctuations via the following distributions of the (classical) boson variables
\begin{align}
    \langle b_{a,i} \rangle = \langle b_{a,i}^\ast \rangle = 0 \quad, \quad \langle b_{a,i}^\ast b_{a,i} \rangle = \tilde{N}_{a,i} +\frac{1}{2} ,
\end{align}
where $\tilde{N}_{a,i}$ are given by $\tilde{N}_{\mathrm{Na},1} = N_{\mathrm{Na}}$, $\tilde{N}_{\mathrm{Li},1} = N_{\mathrm{Li}}$, and $\tilde{N}_{\mathrm{Na},0} = N_{\mathrm{Li},0}=0$, with $N_{a}$ being the total atom numbers in each realization for the two species.
Subsequently, we simulate the initial state preparation with the (classical) Hamiltonian
\begin{align}
    H_\mathrm{prep} = -i\Omega (b^\ast_{\mathrm{Na},0}b_{\mathrm{Na},1} -c.c.),
\end{align}
where $c.c.$ denotes complex conjugation. To prepare the initial state, we apply the Hamiltonian $H_\mathrm{prep}$ for a time duration of $t_\mathrm{prep}/\hbar = 2\pi \bar{\eta}_\mathrm{Na}(t_0)/\Omega $.

For the curves shown in \Fig{fig:Dynamics} and \Fig{fig:Imbalance}, we average the dynamics over 100 realizations of the above procedure. The sketches in \Fig{fig:Sketch} were plotted for the same parameters but with mean value and standard deviation for the atom numbers taken from the data of \Fig{fig:Imbalance}. They illustrate the imbalance values $\bar{\eta}_\mathrm{Na}(t_0)=0.15,0.2,0.5$ from left to right.

\subsection{Thermal ensemble}
We compare the data against a thermal ensemble with temperature $T=630(20)$nK, as extracted from TOF-measurements. Specifically, we compute observables $\hat{\mathcal{O}}$ in the canonical ensemble as
\begin{align}
    \langle \hat{\mathcal{O}} \rangle = \frac{1}{Z} \mathrm{Tr} \left(\hat{\mathcal{O}} \hat{P}_M e^{-\beta\hat{H}}\right) ,
\end{align}
where $Z =\mathrm{Tr} \left(\hat{P}_M \mathrm{exp}(-\beta\hat{H})\right) $ is the canonical partition sum. Here, we focus on the gauge-invariant ensemble with fixed total magnetization $\hat{M}$. As our system evolves at high energies, we consider the classical thermal limit in the following, which is valid for high temperatures. We compute the classical partition function as
\begin{align}
    Z_{cl} = \Big[\prod_{a,m_F}  \int\mathrm{d} b_{a,m_F}\mathrm{d} b^\ast_{a,m_F}\Big] e^{-\beta H(\{b_{a,m_F}\})},
\end{align}
where $H$ is the Wigner-Weyl transform of the system Hamiltonian $\hat{H}_\mathrm{NaLi}$.
For fixed total atom numbers and magnetization, we consider the following Hamiltonian in the classical limit 
\begin{align}
H&= \chi L^2_{\mathrm{Na},z} + \Delta L_{\mathrm{Li},z} \nonumber\\
&+ 2\lambda (L_{\mathrm{Na},x}L_{\mathrm{Li},x}+L_{\mathrm{Na},y}L_{\mathrm{Li},y}),
\end{align}
where $L_{a,\pm} = L_{a,x}\pm i L_{a,y}$. The integral can be transformed to the coordinates
\begin{align}
    Z_{cl} = \mathcal{N}\int \mathrm{d}s \mathrm{d}\phi e^{-\beta H(s,\phi)} ,
\end{align}
where $\mathcal{N}$ is an (unimportant) normalization, and $s$, $\phi$ are the two remaining degrees of freedom of the classical system after using the conservation of magnetization, particle number and integrating out irrelevant absolute phases. Besides the thermal fluctuations represented by this partition sum, we average the results over $20$ realizations of Gaussian atom number fluctuations for Na and Li with $N_\mathrm{Na} = (313.2\pm 18.3)\times 10^3$, $N_\mathrm{Li} = (34.6\pm 3.2)\times 10^3$ for \Fig{fig:Dynamics} and $N_\mathrm{Na} = (376.5\pm 20.4)\times 10^3$, $N_\mathrm{Li} = (29.0\pm 2.3)\times 10^3$ for \Fig{fig:Imbalance}, where the values are mean and standard deviation of the measured overall atom number distributions.

Physically, $\phi$ represents the relative azimuthal angle between both spins and $s = 2\eta_\mathrm{Li}-1$ is the normalized $z$ component of lithium. In this representation, the Hamiltonian is given by
\begin{align}
\label{eq:Ham-Thermo}
    H&(s,\phi) = \chi \ell_\mathrm{Li}^2 s^2 + (\Delta -2M\chi) \ell_\mathrm{Li} s\nonumber\\
    &+2 \lambda \ell_\mathrm{Li}\ell_\mathrm{Na} \sqrt{1-s^2} \sqrt{1- \frac{(M-\ell_\mathrm{Li} s)^2}{\ell_\mathrm{Na}^2}}\cos(\phi) ,
\end{align}
where we defined the spin lengths $\ell_a = N_a/2$. To derive Eq.~\eqref{eq:Ham-Thermo}, we used
\begin{align}
L_{a,x} &= \sqrt{N_{a,0}N_{a,1}}\cos(\phi_a)\\
L_{a,y} &= \sqrt{N_{a,0}N_{a,1}}\sin(\phi_a)\\
L_{a,z} &= \frac{N_{a,0}-N_{a,1}}{2}  
\end{align}
where $\phi_a$ are the relative phases between the magnetic substates for each species. We then set $s_a = (N_{a,0}-N_{a,1})/(N_{a,0}+N_{a,1})$, and used
\begin{align}
    N_{0,a}N_{1,a}&= \left(\frac{N_{a}}{2}+L_{z,a}\right)\left(\frac{N_{a}}{2}-L_{z,a}\right)\\
    &= \left(\frac{N_{a}}{2}\right)^2-L^2_{z,a}\\
    &= \ell_a^2(1-s_a^2),
\end{align}
which allowed us to rewrite the interaction term as
\begin{align}
    L_{\mathrm{Na},x}&L_{\mathrm{Li},x}+L_{\mathrm{Na},y}L_{\mathrm{Li},y} \nonumber\\
    &= \ell_\mathrm{Li}\ell_\mathrm{Na} \sqrt{1-s_\mathrm{Na}^2}\sqrt{1-s_\mathrm{Li}^2}\nonumber\\
    &\quad \times \left(\cos(\phi_\mathrm{Na})\cos(\phi_\mathrm{Li})+\sin(\phi_\mathrm{Na})\sin(\phi_\mathrm{Li})\right)\nonumber\\
    &= \ell_\mathrm{Li}\ell_\mathrm{Na} \sqrt{1-s_\mathrm{Na}^2}\sqrt{1-s^2}\cos(\phi_\mathrm{Na}-\phi_\mathrm{Li}) \; .
\end{align}
Eq.~\eqref{eq:Ham-Thermo} is obtained by further using $M =\ell_\mathrm{Li} s + \ell_\mathrm{Na} s_\mathrm{Na}$ and defining the relative phase $\phi = \phi_\mathrm{Na}-\phi_\mathrm{Li}$ while integrating out the absolute phase $\phi_\mathrm{Na}+\phi_\mathrm{Li}$.

Observables are subsequently obtained by numerical integration of the partition sum. For instance, for the expectation value of the lithium $z$-spin, we get
\begin{align}
    \langle L_{\mathrm{Li},z} \rangle = \int \mathrm{d}s \mathrm{d}\phi\, \ell_{\mathrm{Li}} s e^{-\beta H(s,\phi)} \, ,
\end{align}
and its fluctuations are computed as 
\begin{align}
    \langle L^2_{\mathrm{Li},z} \rangle = \int \mathrm{d}s \mathrm{d}\phi\, \ell^2_\mathrm{Li} s^2 e^{-\beta H(s,\phi)} \, .
\end{align}
In the main text we show connected moments, which are defined as the corresponding variance
\begin{align}
    \sigma^2_{L_{\mathrm{Li},z}} = \langle L^2_{\mathrm{Li},z} \rangle - \langle L_{\mathrm{Li},z} \rangle^2 .
\end{align}

\subsection{Effective potentials}
\begin{figure}[h!]
\centering
    \includegraphics[width=0.47\textwidth]{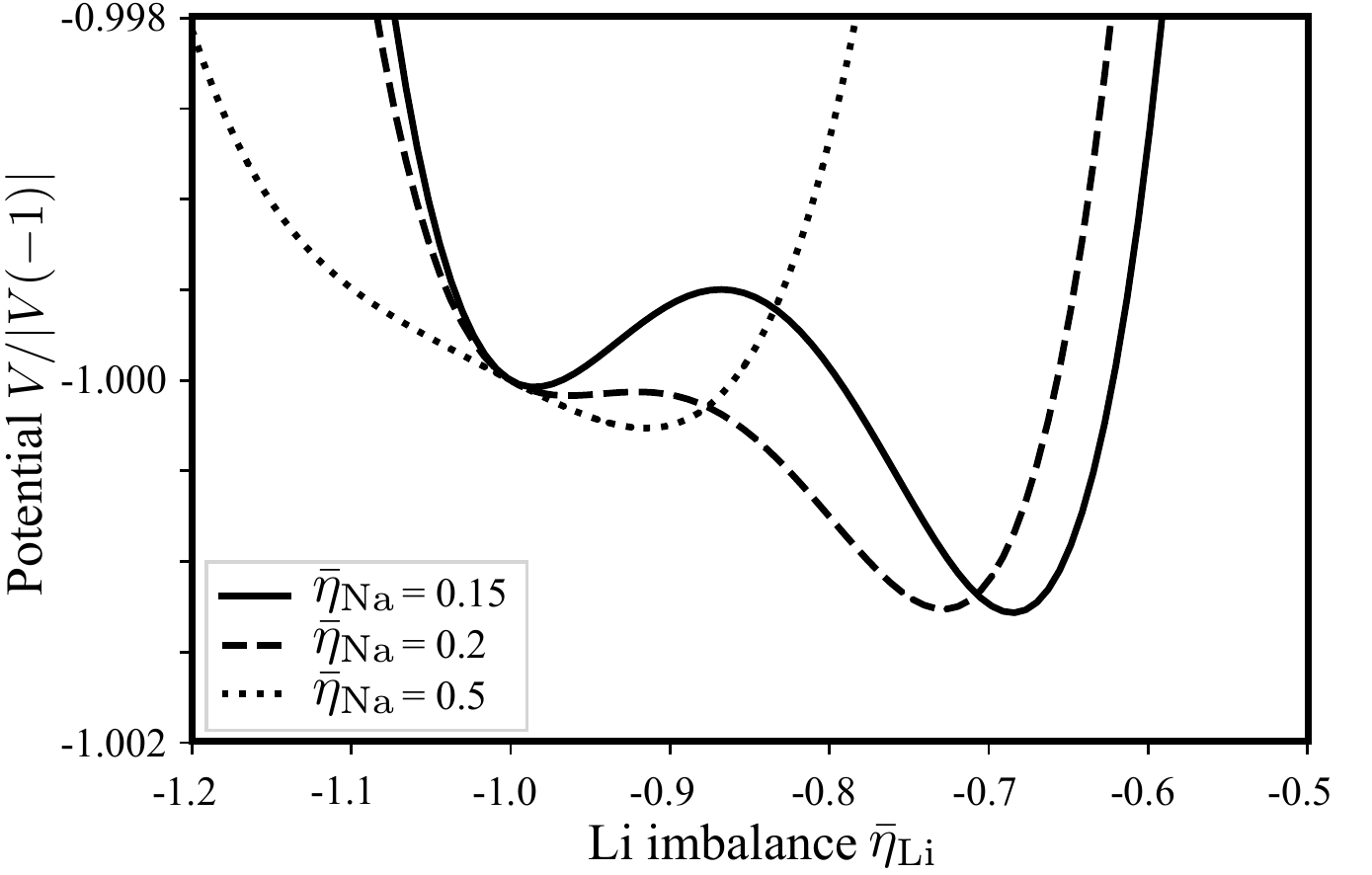}
    \caption{\textbf{Effective mean-field potentials:} We show the effective mean-field potential for $\bar{\eta}_\mathrm{Li}$ for the values $\bar{\eta}_\mathrm{Na}(t_0)=0.15,0.2$, and $0.5$.}
    \label{fig:Potentials}
\end{figure}

In this section we give a more detailed explanation of the effective potentials for mean fields as shown in the main text, see also \cite{torsten2020}. Starting from the model Hamiltonian, the coherent evolution equation relevant for early times are given by the classical equations of motion $\partial_t  \mathbf{L}_{\mathrm{Li}} = \{H,\mathbf{L}_{\mathrm{Li}} \}$, where $\{\cdot,\cdot\}$ denote the fundamental Poisson brackets with $\{ L_{a,i}, L_{b,j}\} = \epsilon_{ijk}\delta_{ab} L_{a,k}$. Using the conservation of energy ($H$), magnetization ($M$), and total atom numbers ($N_\mathrm{Na}$, $N_\mathrm{Li}$) the evolution may be expressed as
\begin{align}
    \partial_t^2 L_{\mathrm{Li},z} = - \frac{\partial V(L_{\mathrm{Li},z})}{\partial L_{\mathrm{Li},z}} ,
\end{align}
where $V$ is a quartic potential $V(x) = \sum_{i=1}^4 c_i x^{i}/i$ and the coefficients $c_i$ are given by
\begin{align}
    c_1 &= (2\chi M-\Delta)(H-\chi M^2) - \lambda^2 M N_\mathrm{Li},\\
    c_2 &= (2\chi M-\Delta )^2 + 2\chi (\chi M^2 -H)\nonumber\\
    &\quad + \lambda^2 (N_\mathrm{Na}^2 + N_\mathrm{Li}^2 -4M^2 ),\\
    c_3  &= -3 \chi (2\chi M -\Delta) + 12 \lambda^2 M ,\\
    c_4 &= 2\chi^2 -8 \lambda^2 .
\end{align}
Straightforward rescaling with the total atom number of Li yields the equations in the main text. In \Fig{fig:Potentials} we show the potentials $V(x)/V(-1)$ for the imbalance values $\bar{\eta}_\mathrm{Na}(t_0)=0.15,0.2,0.5$. In \Fig{fig:Sketch} these potentials are drawn (from left to right) with arbitrary units on the y-axis.

\clearpage

\end{document}